\documentclass[a4paper,12pt]{article}
\usepackage{amsmath}
\usepackage{amssymb}
\usepackage{amsfonts}
\usepackage{mathrsfs}
\usepackage{cite}
\usepackage[font=small,labelfont=bf]{caption}
\usepackage{graphicx}
\usepackage{lscape}
\usepackage{colordvi}
\usepackage{array}

\makeatletter

\newcommand*\wthelper[2]{%
        \hbox{\dimen@\accentfontxheight#1%
                \accentfontxheight#11.3\dimen@
                $\m@th#1\widetilde{#2}$%
                \accentfontxheight#1\dimen@
        }%
}

\newcommand*\accentfontxheight[1]{%
        \fontdimen5\ifx#1\displaystyle
                \textfont
        \else\ifx#1\textstyle
                \textfont
        \else\ifx#1\scriptstyle
                \scriptfont
        \else
                \scriptscriptfont
        \fi\fi\fi3
}
\makeatother

\def\be{\begin{equation}}
\def\ee{\end{equation}}
\def\beq{\begin{equation}}
\def\eeq{\end{equation}}
\def\bea{\begin{eqnarray}}
\def\eea{\end{eqnarray}}

\def\simgt{\stackrel{>}{{}_\sim}}

\def\<{\left\langle}
\def\>{\right\rangle}

\newcommand{\vev}[1]{\left\langle{#1}\right\rangle}

\begin{document}

\bibliographystyle{OurBibTeX}

\title{\hfill ~\\[-30mm]
       \hfill\mbox{\small }\\[13mm]
       \sffamily\huge{A Golden $A_5$ Model of Leptons\\ with a Minimal NLO Correction}}

\author{
Iain~K.~Cooper\footnote{E-mail: \texttt{ikc1g08@soton.ac.uk}}\:,~~
Stephen~F.~King\footnote{E-mail: \texttt{king@soton.ac.uk}}\\~~and~~
Alexander~J.~Stuart\footnote{E-mail: \texttt{a.stuart@soton.ac.uk}}\\
\\[9mm]
{\small\it
School of Physics and Astronomy,
University of Southampton,}\\
{\small\it Southampton, SO17 1BJ, U.K.}\\[2mm]
}

\date{}

\maketitle
\thispagestyle{empty}

\begin{abstract}
\noindent
We propose a new $A_5$ model of leptons which corrects the LO predictions
of Golden Ratio mixing via a minimal NLO Majorana mass correction which completely breaks the original
Klein symmetry of the neutrino mass matrix. The minimal
nature of the NLO correction leads to a restricted and correlated range of the mixing angles
allowing agreement within the one sigma range of recent global fits following the reactor angle measurement by Daya Bay and RENO.
The minimal NLO correction also preserves the LO inverse neutrino mass sum rule leading to a neutrino mass spectrum that extends into the quasi-degenerate region allowing the model to be 
accessible to the current and future neutrinoless double beta decay experiments.
\end{abstract}

\newpage
\setcounter{footnote}{0}

\section{Introduction}
\label{sec:introduction}
It is an interesting feature of neutrino physics that two of the lepton mixing angles, the atmospheric
angle $\theta_{23}$ and the solar angle $\theta_{12}$, are both 
rather large \cite{pdg}.  Until recently the remaining reactor
angle, $\theta_{13}$, was unmeasured. Direct evidence for $\theta_{13}$ was
first provided by T2K, MINOS and Double
Chooz~\cite{Abe:2011sj,Adamson:2011qu,Abe:2011fz}. 
Recently Daya Bay~\cite{DayaBay}, RENO~\cite{RENO}, and Double
Chooz~\cite{DCt13} Collaborations have measured $\sin^2(2\theta_{13})$:
\begin{eqnarray}
\label{t13}
\begin{array}{cc}
\text{Daya Bay: } & \sin^2(2\theta_{13})=0.089\pm0.011 \text{(stat.)}\pm0.005
\text{(syst.)}\ ,\\
\text{RENO: }    & \sin^2(2\theta_{13})= 0.113\pm0.013\text{(stat.)}\pm0.019
\text{(syst.)\ ,}\\
\text{Double Chooz: }& \sin^2(2\theta_{13})=0.109\pm0.030\text{(stat.)}\pm0.025
\text{(syst.)}\ .\\
\end{array}
\end{eqnarray}

This measurement excludes the tri-bimaximal (TB) lepton mixing
pattern\cite{TBM} in which the atmospheric angle is maximal, the reactor angle
vanishes, and the solar mixing angle is approximately $35.3^{\circ}$. 
When comparing global fits to TB mixing it is convenient to express the solar,
atmospheric and reactor angles in terms of deviation parameters ($s$, $a$ and
$r$) from TB mixing\cite{rsa}: 
\be
\label{rsadef}
\sin \theta_{12}=\frac{1}{\sqrt{3}}(1+s),\ \ \ \ 
\sin\theta_{23}=\frac{1}{\sqrt{2}}(1+a), \ \ \ \ 
\sin \theta_{13}=\frac{r}{\sqrt{2}}.
\ee
From the global fits in \cite{global,global2,global3} for illustrative purposes 
one may consider the representative $1\sigma$ ranges for the TB deviation parameters:
\beq
s=-0.03\pm 0.03, \ \ \ \ a=-0.10 \pm 0.03, \ \ \ \ r=0.22\pm 0.01,
\label{rsafit}
\eeq
assuming a normal neutrino mass ordering and a first octant atmospheric mixing angle. Notice that even though these ranges are quoted for the normal hierarchy, the representative ranges for the inverted hierarchy lie within these normal hierarchy ranges when considering a first octant atmospheric mixing angle. 
We emphasise that the global fits do not decisively determine the octant for the atmospheric mixing angle.
As well as showing that TB is excluded by the reactor angle being non-zero, 
Eq.~(\ref{rsafit}) indicates a preference for the atmospheric angle to
be below its maximal value and also a slight preference for the solar angle to
be below its tri-maximal value. It seems that TB mixing no longer holds the exalted
position that it did before, and perhaps now is an opportune moment 
to consider other mixing patterns that have been proposed but which have so far been somewhat eclipsed
by TB mixing.

An interesting alternative to TB mixing is the Golden Ratio (GR) \cite{goldenramond,goldenstrumia,goldeneverett}
mixing pattern:
\begin{equation}
\label{UGR}
 U_{GR}=\left(
\begin{array}{ccc}
 \sqrt{\frac{\phi_g}{\sqrt{5}}} & \sqrt{\frac{1}{\phi_g\sqrt{5}}} & 0 \\
 -\sqrt{\frac{1}{2\phi_g\sqrt{5}}} & \sqrt{\frac{\phi_g}{2\sqrt{5}}} & \frac{1}{\sqrt{2}} \\
 \sqrt{\frac{1}{2\phi_g\sqrt{5}}} & -\sqrt{\frac{\phi_g}{2\sqrt{5}}} & \frac{1}{\sqrt{2}}
\end{array}
\right)P_0,
\end{equation}
so named because it involves $\phi_g=\frac{1+\sqrt{5}}{2}$, the famed Golden Ratio which ancient Greeks thought aesthetically pleasing.  Notice that we have reported the above lepton mixing matrix in the Particle Data Group (PDG) convention for $U_{PMNS}$\cite{pdg} where $P_0=\text{Diag}(1,e^{\frac{i\alpha^0_2}{2}},e^{\frac{i\alpha^0_3}{2}})$ is the matrix of Majorana phases.

The history of the GR's possible role in lepton mixing began as a footnote\cite{goldenramond}.  Several years later, this idea was applied \cite{goldenstrumia} in the context of a non-diagonal charged lepton basis by observing that $\theta_{12}=\tan^{-1}(1/\phi_g)\approx 31.7^{\circ}$ was a good leading order (LO) prediction for the solar neutrino mixing angle.  In addition to this, Ref. \cite{goldenstrumia} conjectured the possible connection of this prediction to the group $A_5$.  With this in mind, the authors of Ref. \cite{goldeneverett} sought to elucidate the group theory of $A_5$ and generate the aforementioned prediction for the solar neutrino angle prediction in the context of a non-dynamical flavour model.\footnote{It is important to note that Ref. \cite{A5first} laid a large portion of the group theoretics of $A_5$ used in the exploration of Ref. \cite{goldeneverett}.}  Shortly after this, it was found \cite{goldenrodejohann} that there was another possible prediction for the solar neutrino mixing angle involving the Golden Ratio, $\theta_{12}=\cos^{-1}(\phi_g/2)=36^{\circ}$.  But instead of using $A_5$, the dihedral group $D_{10}$ was utilised to dynamically generate this prediction.  For two years, the idea of the Golden Ratio's possible implication in neutrino mixing lay seemingly dormant\footnote{With the exception of Ref. \cite{alexlisaquarks} constructing a quark model by extending $A_5$ to its double cover, $A_5^{\prime}$, with their golden ratio prediction of Ref. \cite{goldeneverett} in mind.}, until a dynamical $A_5$ model 
was constructed \cite{goldenparis} in a basis in which the charged lepton mass matrix is diagonal.  Shortly afterwards,  it was shown \cite{goldending} that the prediction of $\theta_{12}=\tan^{-1}(1/\phi_g)\approx 31.7^{\circ}$ was minimally realised in $A_5$, when assuming a diagonal charged lepton basis.  Yet a group in which the prediction $\theta_{12}=\cos^{-1}(\phi_g/2)$ was realised under the same assumptions, could not be found.  The same work\cite{goldending} also contained two dynamical models which predicted Golden Ratio mixing.  It should be noted that as all of this work using $A_5$ to construct models which predicted Golden Ratio lepton mixing was being done, $A_5$ was also being used to construct a four family lepton model\cite{A54gen}, and its double cover, $A_5^{\prime}$, to construct a four family model incorporating quarks\cite{A5prime4gen}.  In addition to this, $A_5^{\prime}$ was used to construct a flavour model explaining cosmic-ray anomalies \cite{A5primedarkmatter}.

Unfortunately the measurement of the reactor angle also excludes the GR
lepton mixing pattern in Eq.(\ref{UGR}) in which the atmospheric angle is maximal, the reactor angle
vanishes, and the solar mixing angle is given by $\theta_{12}=\tan^{-1}(1/\phi_g)\approx 31.7^{\circ}$,
corresponding to $r=a=0$ and $s\approx -0.09$. 
These all lie outside the $1\sigma$ ranges in Eq.(\ref{rsafit}), especially the reactor parameter $r$,
but also the atmospheric deviation parameter $a$, and even the solar parameter $s$ is too negative.
In the face of this disagreement, we shall interpret the GR prediction as the LO prediction and then try to achieve consistency with the data at next-to-leading order (NLO). This is analogous to a recent strategy that was developed for dealing
with the TB prediction in the light of a non-zero reactor angle \cite{trustsym}, where the LO Klein symmetry
based on the $S_4$ generators $S,U$ was broken at NLO but the generator $S$ was preserved in order to 
maintain the successful tri-maximal solar prediction $s=0$, leading to an atmospheric sum rule
relation $a=-(r/2)\cos \delta$.
However, in the present case, since we also wish to correct the solar angle prediction, we 
do not wish to preserve any of the original LO Klein symmetry, and so we shall break both the corresponding $S$ and $U$ Klein subgroup generators of $A_5$ defined in Table~\ref{tab:A5Gen}
at NLO. For a general NLO correction, this would imply arbitrary values for the deviation
parameters $r,s,a$. However we shall consider a minimal NLO correction leading to restricted and
correlated ranges of these parameters. 

In the present paper, then, we propose a new $A_5$ model of leptons which corrects the LO prediction
of GR mixing via a minimal NLO Majorana mass correction that completely breaks the original
Klein symmetry generators, $S$ and $U$, thereby correcting all three mixing angles and
allowing agreement to be achieved with the global fits. Although there is no remaining mixing angle 
sum rule prediction, the minimal
nature of the assumed NLO correction leads to a restricted and correlated range of the resulting deviation parameters
$r,s,a$ which encompasses the $1\sigma$ ranges of these parameters from the recent global fits.
The minimal NLO Majorana mass correction also
has the feature that it preserves a LO inverse neutrino mass
sum rule even at NLO, leading to a quasi-degenerate neutrino mass spectrum.\footnote{Note that if a singlet flavon is added then the sum rule does not exist at LO.}
This, in turn, severely constrains the parameter space of neutrinoless double beta decay, allowing the model to be tested by next generation neutrinoless double beta decay experiments.

The layout of the remainder of the paper is as follows. In Section \ref{sec:model}, the Golden Lepton Flavour Model is constructed by defining its fields and transformation properties under $A_5$ as well as an additional $U(1)$ symmetry.  The resulting LO and NLO mass matrices are constructed after electroweak and flavour symmetry breaking and diagonalised to reveal the lepton mass and mixing parameters.   In Section \ref{sec:alignment}, the vacuum alignment of the model is analysed by explicit construction and minimisation of the model's flavon potential.  Section \ref{sec:sumruleimp} contains a detailed analysis of the phenomenological implications of the NLO correction on the TB deviations parameters ($r$, $s$, and $a$) and the effective Majorana mass scale of neutrinoless double beta decay.  In Section \ref{sec:conclusion}, the discussion of the $A_5\times U(1)$ model is concluded.  The relevant group theory of $A_5$ can be found in Appendix \ref{sec:appendixa}
and the breaking of the low-energy Golden Ratio Klein symmetry is discussed in Appendix \ref{sec:appendixc}.

\section{The Model }
\label{sec:model}
\subsection{Fields, Symmetries, and Yukawa Superpotentials}
In this section, we present an $A_5$ model of leptons as a framework in which to study the generation of a non-zero $\theta_{13}$ from $A_5$.  We begin this discussion by noting that the left-handed lepton doublets $L=(l_e,l_{\mu},l_{\tau}$), the right-handed charged lepton singlets $E=(e_R^c,\mu_R^c,\tau_R^c)$, and the right-handed neutrinos $N=((\nu^c_R)_e,(\nu^c_R)_{\mu},(\nu^c_R)_{\tau})$ all transform under the $\bf{3}$-dimensional irreducible representation of $A_5$.  Furthermore, the up- and down-type Higgs doublet fields $h_u$ and $h_d$
are assumed to be blind to the flavour symmetry.  As a result of this minimal field content and its transformation properties $h_uLN$, $NN$, and $h_dLE$ are the only allowed terms in the Yukawa superpotential because $\bf{3\otimes 3=1_s\oplus 3_a \oplus 5_s}$ contains a singlet (see Appendix \ref{sec:appendixa} for the group theory of $A_5$).  However, these terms lead to undesirable
phenomenological predictions, namely degenerate charged lepton and neutrino masses as well as no leptonic mixing.

To fix the problematic predictions of the simplest LO terms ($h_uLN$, $NN$, and $h_dLE$), it is necessary to introduce additional scalar fields (\textit{i.e.} flavon fields) which will couple to the existing matter fields, as well as an additional $U(1)$ symmetry to forbid problematic operators which lead to un-phenomenological results.  The $U(1)$ symmetry will not be gauged, in order to avoid constraints associated with anomaly cancellation.  Yet, it will be spontaneously broken by the flavon fields acquiring vacuum expectation values (VEVs).  In general, this will lead to massless Goldstone bosons, unless the symmetry is also explicitly broken.  Therefore, the $U(1)$ symmetry is assumed to be explicitly broken in the hidden sector of the theory, so that that Goldstone bosons become Pseudo-Goldstone bosons with mass around 1 TeV.  The additional fields, as well as the original matter fields, and their transformation properties under $A_5\times U(1)$ can be found in Table \ref{tab:matter}.  Using these transformation properties, the Yukawa operators invariant under the $A_5\times U(1)$ symmetry can be constructed.  
\begin{table}[h]
	\centering
		\begin{tabular}{|c||c|c|c||c|c|c|c|c|c|}
			\hline
				Field & $L$ & $E$ & $N$ & $h_{u,d}$ & $\chi$ & $\phi$ & $\theta$ & $\lambda$ & $\varphi$ \\ \hline \hline
				$A_5$ & ${\bf 3}$ & ${\bf 3}$ & ${\bf 3}$ & ${\bf 1}$ & ${\bf 5}$ & ${\bf 5}$ & ${\bf 1}$ &  ${\bf 3}$ & ${\bf 5}$ \\ \hline
				$U(1)$ & $-y$ & $5x+y$ & $y$ & $0$ & $-5x$ & $-3x$ & $-x$ & $-y$ & $-2y$  \\ \hline
		\end{tabular}
	\caption{The field content of the $A_5\times U(1)$ model where $x$ and $y$ are carefully chosen integers.
	For example we have checked that $x=-9$ and $y=2$ leads to the desired operators with no undesirable operators. }
\label{tab:matter}
\end{table}

The construction of the $A_5\times U(1)$ invariant Yukawa superpotential is begun in the neutrino sector.  Using the fields and transformation properties defined in Table \ref{tab:matter}, the neutrino Yukawa superpotential may be schematically written to NLO as
\begin{equation}
 W_{\nu}=yLNh_u+y_2 NN\varphi+\Delta W_{\nu},
\label{eqn:nuterms}
\end{equation}
where $\Delta W_{\nu}$ denotes the NLO correction to the superpotential which will generate a nonzero $\theta_{13}$ and a correction to $\theta_{23}$ and $\theta_{12}$; it will be discussed more fully in Section \ref{sec:theta13}.  Proceeding to the charged lepton sector, the superpotential resulting in the generation of the charged lepton masses and mixings can be schematically expressed as
\begin{equation}
 W_l=\frac{y_4}{\Lambda} LE \chi h_d+\frac{y_5}{\Lambda^3}LE \theta^2\phi  h_d+\frac{y_6}{\Lambda^5}LE\theta^5h_d.
\label{eqn:leptonterms}
\end{equation}
With the charged lepton and neutrino superpotentials in hand, we now turn to the 
construction of the charged lepton and neutrino mass matrices after electroweak and flavour symmetry breaking.

\subsection{The Charged Lepton Mass Matrix and its Predictions}
\begin{table}[t]
	\centering
		\begin{tabular}{|c|c|}
			\hline
				Flavon VEV & VEV alignment\\ \hline
				$\left\langle\phi\right\rangle^{\phantom{T}}$ & $(0,0,0,v_{\phi},0)^T$ \\ \hline
				$\left\langle\chi\right\rangle^{\phantom{T}}$ & $(0,0,v_{\chi},0,0)^T$ \\ \hline
				$\left\langle\lambda\right\rangle^{\phantom{T}}$ & $(0,0,a_3)^T$ \\ \hline
				$\left\langle\varphi\right\rangle^{\phantom{T}}$ & $(\sqrt{\frac{2}{3}}\left(v_2+v_3\right),v_2,v_3,v_3,v_2)^T$ \\ \hline
		\end{tabular}
	\caption{The vacuum alignments of the flavons used in
          the model in terms of complex VEVs. Justification of the flavon alignments follows from a detailed discussion of the minimisation of the flavon potential in Section \ref{sec:alignment}.}
	\label{tab:flavonalignments}
\end{table}

Using the superpotential in Eq. (\ref{eqn:leptonterms}), the charged lepton mass matrix, after electroweak and flavour symmetry breaking, can be constructed.  This is done by utilising the Clebsch-Gordan Coefficients of Appendix \ref{sec:appendixa} to appropriately contract the flavon fields, thereby extracting the singlets from the product representations.  Then, the VEVs of 
$\phi$ and $\chi$ given in Table~\ref{tab:flavonalignments}, as well as $\langle\theta\rangle=v_{\theta}$ are applied to reveal the charged lepton mass matrix, $M_e$, after electroweak and flavour symmetry breaking, to be
\begin{equation}
 M_e=v_d\begin{pmatrix}
      \frac{y_6}{\Lambda^5}v_{\theta}^5 & 0 & 0 \\
      0 & \sqrt{6}\frac{y_5}{\Lambda^3}v_{\theta}^2v_{\phi} & \frac{y_6}{\Lambda^5}v_{\theta}^5 \\
      0 & \frac{y_6}{\Lambda^5}v_{\theta}^5 & \sqrt{6}\frac{y_4}{\Lambda}v_{\chi}
     \end{pmatrix}.
\label{eqn:leptonmass}
\end{equation}
By assuming all flavon field VEVs are of the same order, \textit{i.e.} $\frac{v_i}{\Lambda}\sim\epsilon$, and absorbing factors of $\sqrt{6}$ into $y_4$ and $y_5$, a charged lepton hierarchy of $\epsilon^4:\epsilon^2:1$ is obtained.  With the added assumption that $\epsilon\sim 0.15$, a phenomenologically viable charged lepton mass spectrum is obtained.  Furthermore, the associated charged lepton mixing matrix is the identity to $\mathcal{O}(\epsilon^4)$.  Therefore, charged lepton mixing may be neglected with respect to the mixing originating from the neutrino sector, as will be seen.


\subsection{The Leading Order Neutrino Mass Matrix}
\label{sec:LOModel}


In the case of $A_5$ a non-zero value for $\theta_{13}$ can be generated by adding flavons which develop VEVs which break the $U$ generator of the Klein symmetry (cf. similar logic was applied in Ref. \cite{trustsym}). Yet before adding these flavons whose VEVs break the Klein symmetry leading to deviations away from GR mixing, we must deduce which flavons should be added whose VEVs preserve the Klein symmetry and lead to GR mixing. 

To deduce the possible irreducible representations of the Klein symmetry
preserving flavons, we begin by noting that since we choose to couple all of our flavons to $NN$, where $N\sim {\bf 3}$ and ${\bf 3}\otimes{\bf 3}={\bf 1_s}\oplus{\bf 3_a}\oplus{\bf 5_s}$, three flavons could be added transforming under the corresponding $\bf{1_s}$, $\bf{3_a}$, and $\bf{5_s}$ representations.  Yet the $\bf{3_a}$ contained in ${\bf 3}\otimes{\bf 3}$ is antisymmetric (as denoted by the subscript ``$\bf{a}$''), and hence it vanishes.  One could include the $\bf{1_s}$ from the ${\bf 3}\otimes{\bf 3}$, as in Ref. \cite{goldending}, but its addition alone will not yield GR mixing, and when added with the $\bf{5_s}$ in ${\bf 3}\otimes{\bf 3}$ will not lead to a LO neutrino mass sum rule.  Thus we choose not to include it.
Therefore, to LO we only include a single flavon $\varphi\sim 5$ which aquires a VEV (up to basis transformations) $\left\langle\varphi\right\rangle^{\phantom{T}}=(\sqrt{\frac{2}{3}}\left(v_2+v_3\right),v_2,v_3,v_3,v_2)^T$,
where $v_2$ and $v_3$ are left unspecified by $A_5$.
This VEV alignment is preserved by the 
five dimensional representations of the $S$ \textit{and} $U$ generators in Appendix~\ref{sec:appendixa},
and hence respects the corresponding $Z_2^S\times Z_2^U$ Klein symmetry
of the LO neutrino Majorana mass matrix,
\begin{eqnarray}
M^{LO}_{R}=y_2\left(
\begin{array}{ccc}
      2\sqrt{\frac{2}{3}}\left(v_2+v_3\right) & -\sqrt{3}v_2 & -\sqrt{3}v_2 \\
      -\sqrt{3}v_2 & \sqrt{6}v_3 & -\sqrt{\frac{2}{3}}\left(v_2+v_3\right)  \\
      -\sqrt{3}v_2 & -\sqrt{\frac{2}{3}}\left(v_2+v_3\right) &  \sqrt{6}v_3
\end{array}\right).
\label{eqn:majoranamass}
\end{eqnarray}
It is straightforward to see that
\begin{equation}
U_{GR}^TM^{LO}_{R}U_{GR}=\text{Diag}(M_1^{LO},M_2^{LO},M_3^{LO}),
\end{equation}
where
\begin{eqnarray}
\begin{array}{ccc}
M_1^{LO}=\frac{y_2 \left(v_2 (6 \phi_g -2)+4 v_3\right)}{\sqrt{6}},&M_2^{LO}=\frac{y_2 \left(4 v_3-v_2 \left(\frac{6}{\phi_g }+2\right)\right)}{\sqrt{6}},&M_3^{LO}=\frac{y_2\sqrt{2}\left(v_2+4 v_3\right) }{\sqrt{3} }.
\end{array}
\end{eqnarray}
It is clear the above masses obey the sum rule
\begin{equation}
\label{majsumrule}
M_1^{LO}+M_2^{LO}=M_3^{LO}.
\end{equation}
Having constructed and diagonalised $M^{LO}_{R}$, it is trivial to construct the LO Dirac mass matrix, as it is just the $\bf{1_s}$ in ${\bf 3}\otimes{\bf 3}$ resulting from the $yLNh_u$ operator in the superpotential (cf. Eq. (\ref{eqn:nuterms})).  After the spontaneous breaking of electroweak and flavour symmetries, the LO Dirac mass matrix has the form
\begin{equation}
 m_D=y v_u\begin{pmatrix}
 	       1 & 0 & 0 \\
	       0 & 0 & 1 \\
	       0 & 1 & 0
 	      \end{pmatrix}.
\label{eqn:diracmass}
\end{equation}
Notice that the LO neutrino mass matrices given in Eqs. (\ref{eqn:majoranamass}) and (\ref{eqn:diracmass}) respect Form Dominance \cite{FD}.
Applying the Seesaw Mechanism\cite{seesaw} to $m_D$ and $M^{LO}_{R}$ generates the light neutrino mass matrix, $M_{\nu}^{LO}$,
that is diagonalised by the $U_{GR}$ of. Eq. (\ref{UGR}) after a matrix of unphysical phases $P^{\prime}=\text{Diag}(1,1,-1)$ has been applied to $U_{GR}$.  Furthermore, we will neglect the Majorana phases of $U_{GR}$ in the diagonalisation so the resulting masses are complex.  Performing this procedure yields the complex light neutrino masses 
\begin{eqnarray}
\label{LOneutmass}
\begin{array}{ccc}
m_1^{LO}=\frac{\sqrt{6} y^2 v_u^2}{y_2 \left(v_2 (6 \phi_g -2)+4 v_3\right)},&m_2^{LO}=\frac{\sqrt{6} y^2 v_u^2}{y_2 \left(4 v_3-v_2 \left(\frac{6}{\phi_g }+2\right)\right)},&m_3^{LO}=\frac{\sqrt{\frac{3}{2}} y^2 v_u^2}{\left(v_2+4 v_3\right) y_2}.
\end{array}
\end{eqnarray}
Notice that these complex masses obey the inverse neutrino mass sum rule \cite{goldending}
\begin{equation}
\label{sumrule}
\frac{1}{m_1^{LO}}+\frac{1}{m_2^{LO}}=\frac{1}{m_3^{LO}}.
\end{equation}

For the remainder of this work, it is useful to re-express the complex neutrino masses in terms of new parameters $\beta$, $\xi$, and $\delta$ such that 
\begin{eqnarray}
\label{redmass}
\begin{array}{ccc}
m_1^{LO}=\frac{ \beta }{6\phi_g-2+4 e^{i \delta } \xi},& m_2^{LO}=\frac{ \beta }{4 e^{i \delta } \xi-(\frac{6}{\phi_g} +2) }, &m_3^{LO}=\frac{ \beta }{2 \left(1+4 e^{i \delta } \xi \right)}.
\end{array}
\end{eqnarray}
To arrive at the above forms of the complex neutrino masses, let $v_2=|v_2|e^{i\theta_2}$ and $v_3=|v_3|e^{i\theta_3}$.  Then, define $\xi=|v_3|/|v_2|$ and $\delta=\theta_3-\theta_2$.  Then, it is clear $\beta=\frac{v_u^2 y^2\sqrt{6}}{y_2 v_2}$.  Notice that $\xi$ and $\delta$ are real parameters,
where $\delta$ should not be confused with the Dirac CP phase.  
Note that the argument of $\beta$ corresponds to an overall phase and hence is unphysical.

The re-parametrisation of the complex neutrino masses in terms $\beta$, $\xi$, and $\delta$ concludes the construction of the LO lepton model.  Notice that this model yields a zero reactor angle,
maximal atmospheric mixing, and a solar mixing angle given by $\theta_{12}=\tan^{-1}(1/\phi_g)\approx 31.7^{\circ}$
which are all in conflict recent global fits.  Therefore, it is necessary to consider
the NLO corrections to the existing minimal LO GR model resulting from the addition of operators which, upon application of the flavon fields' VEVs, lead to corrections to the neutrino mass matrix that leave it no longer invariant under the $S$ and $U$ generators, providing a correction to the problematic leading order predictions.


\subsection{The Neutrino Sector at NLO}
\label{sec:theta13}


In order to generate a non-zero reactor mixing angle in the context of $A_5$, a flavon field must be added which breaks the $U$ generator of the $Z_2^S\times Z_2^U$ Klein symmetry.  To do this, an additional flavon field $\lambda$ transforming as a ${\bf 3}$ under $A_5$ is introduced.  Recall that the contraction of a flavon field transforming as a $\bf{3}$   and $NN$ necessarily vanishes due to the anti-symmetry of the ${\bf 3_a}$ contained in ${\bf 3\otimes 3}$.  Thus, we allow for a quadratic coupling of $\lambda$ to $NN$.  As a result, the superpotential receives a next-to-leading-order correction of
\footnote{The pair of contractions given here form a basis for all possible contractions.}, 
\begin{equation}
 \Delta W_{\nu}=\frac{y_{A}}{\Lambda}\left(NN\right)_{\bf 1_s}\left(\lambda\lambda\right)_{\bf 1_s}+\frac{y_{C}}{\Lambda}\left(\left(NN\right)_{\bf 5_s}\left(\lambda\lambda\right)_{\bf 5_s}\right)_{\bf 1_s}.
\label{eqn:newMaj}
\end{equation}
This corrects the LO Majorana mass matrix in Eq. \eqref{eqn:majoranamass} to yield
\begin{equation}
 M_{R}=M_{R}^{LO}+\Delta M_{R},
\label{eqn:majtot}
\end{equation}
where $M_{R}^{LO}$ is defined in Eq. \eqref{eqn:majoranamass}.  Before proceeding to calculate $\Delta M_{R}$, it is useful to note that the coefficient $y_A$ will not enter the eigenvectors of the corrected/perturbed mass matrix
since the associated operator has a simple structure in which
its contribution to $\Delta M_{R}$ is of the form of a singlet of $A_5$ which is compatible with the preservation the Golden Ratio Klein Group.  Thus, such a contribution still leads to GR mixing. 
Therefore, it will not affect the corrected mixing.\footnote{However, its existence will shift the neutrino masses depending on the alignment of $\langle\lambda\rangle$.}  Thus, we choose an alignment for $\langle\lambda\rangle=(a_1,a_2,a_3)^T$ such that the operator associated with $y_A$ vanishes. This will require $a_1=0$ and either $a_2$ or $a_3$ to be zero.  We choose $a_2=0$ implying $\langle\lambda\rangle=(0,0,a_3)^T$.  This choice will be justified in the detailed discussion of the minimisation of the flavon potential in Section \ref{sec:alignment}.  

Having assumed $\lambda$ develops VEV $\langle\lambda\rangle=(0,0,a_3)^T$, the correction $\Delta M_{R}$ takes the rather simple form
\begin{eqnarray}
\label{eq:majcorr}
\Delta M_{R}=\left(
\begin{array}{ccc}
 0 & 0 & 0 \\
 0 & 6 y_Cv_4 & 0 \\
 0 & 0 & 0
\end{array}
\right),
\end{eqnarray}
where $v_4\equiv a_3^2/\Lambda$.
As discussed in Appendix~\ref{sec:appendixc}, this correction turns out to be sufficient to break \textit{both} the $S$ and $U$ Klein
symmetry generators associated with the unsuccessful LO Golden Ratio predictions of a vanishing reactor angle and a solar angle that which is too small.  Both of these generators must be broken because (from Section \ref{sec:introduction}) breaking $U$ will lead to a nonzero reactor angle and a non-maximal atmospheric angle and breaking $S$ will affect the solar mixing angle prediction. 

Recall that the LO Golden Ratio mixing matrix is given by the usual GR form
as in Eq.~(\ref{UGR}), 
up to charged lepton corrections (which are small as can be seen in Eq. \eqref{eqn:leptonmass}), where $U_{GR}$ has been brought into the PDG convention\cite{pdg}.  
In the presence of $\lambda$, the PMNS mixing matrix is given as
\begin{equation}
 U_{PMNS}'=U_{GR}+\Delta U,
\label{eqn:gr13}
\end{equation}
where the prime on $U_{PMNS}'$ indicates that it is not yet in standard PDG form due to the correction. 
Assuming 
\begin{equation}
v_4\equiv \frac{a_3^2}{\Lambda} \ll v_2, v_3
\end{equation}
enables a perturbative calculation of the correction $\Delta U$.  
To first order in $v_4/v_{2,3}$, the corrections to the columns of the PMNS matrix are
\begin{align}
\label{deltaUbegin}
\Delta U_{11}&\approx-\frac{\epsilon}{5}\sqrt{\frac{3}{2\phi_g\sqrt{5}}},\\ 
\Delta U_{12}&\approx \frac{\epsilon}{5}\sqrt{\frac{3\phi_g}{ 2\sqrt{5} }},\\
\Delta U_{13}&\approx 18\epsilon\sqrt{3}D_1^{-1},\\
\Delta U_{21}&\approx \epsilon\sqrt{3\phi_g}(2\phi_g\xi e^{i\delta}-\phi_g+12)D_2^{-1},\\
\Delta U_{22}&\approx \epsilon\sqrt{\frac{3}{\phi _g}}(12\phi_g+1-2\xi e^{i\delta})D_3^{-1}, \\
\Delta U_{23}&\approx 3\epsilon\sqrt{6}(1-2\xi e^{i\delta})D_1^{-1}, \\
\Delta U_{31}&\approx \epsilon\sqrt{3\phi_g}(2\phi_g\xi e^{i\delta}-\phi_g-18)D_2^{-1}, \\
\Delta U_{32}&\approx \epsilon\sqrt{\frac{3}{\phi_g}}(1-18\phi_g-2\xi e^{i\delta})D_3^{-1},\\\label{deltaUend}
\Delta U_{33}&\approx 3\epsilon\sqrt{6}(1-2\xi e^{i\delta})D_1^{-1},
\end{align}
where 
\begin{align}
D_1&= 2 \sqrt{2}(4\xi^2e^{2i\delta}+2\xi e^{i\delta}-11)\\
D_2&= 10\times  5^{1/4}(\phi_g+3-2\phi_g \xi e^{i\delta}) , \\
D_3&=  10\times  5^{1/4}(3\phi_g-1+2\xi e^{i\delta}),~\text{and}\\\label{correctionend}
\epsilon&=\frac{y_Cv_4}{v_2 y_2}.
\end{align}
For simplicity, we will identify the $\epsilon$ of Eq. (\ref{correctionend}) with $\epsilon\sim v_i/\Lambda$ used in Eq. (\ref{eqn:leptonmass}).  This is justified/motivated by assuming that $y_C$ and $y_2$ are of $\mathcal{O}(1)$ and that any VEV over the cut-off scale $\Lambda$ should be of similar size.  Hence, $v_4=a_3^2/\Lambda\sim \epsilon a_3$ and $a_3\sim v_2$.  From this it is clear that the quantity $\epsilon \sim \frac{v_i}{\Lambda}$
defined below Eq. (\ref{eqn:leptonmass}) and that in Eq. (\ref{correctionend}) will be of similar size and perturbatively indistinguishable.
Therefore the correction to the PMNS matrix is 
\begin{equation}
 \Delta U = \begin{pmatrix}
             \Delta U_{i1} & \Delta U_{i2} & \Delta U_{i3}
            \end{pmatrix}.
\label{eqn:simplePMNS}
\end{equation}

As noted in Section \ref{sec:introduction}, a general lepton mixing matrix can be expressed in terms of the TB deviation parameters $r$, $s$, and $a$.  Doing so allows one to write the lepton mixing matrix as\cite{rsa}
\begin{eqnarray}
U_{\mathrm{PMNS}} \approx
\left( \begin{array}{ccc}
{\frac{2}{\sqrt{6}}}(1-\frac{1}{2}s)  & \frac{1}{\sqrt{3}}(1+s) & \frac{1}{\sqrt{2}}re^{-i\delta } \\
-\frac{1}{\sqrt{6}}(1+s-a + re^{i\delta })  & \phantom{-}\frac{1}{\sqrt{3}}(1-\frac{1}{2}s-a- \frac{1}{2}re^{i\delta })
& \frac{1}{\sqrt{2}}(1+a) \\
\phantom{-}\frac{1}{\sqrt{6}}(1+s+a- re^{i\delta })  & -\frac{1}{\sqrt{3}}(1-\frac{1}{2}s+a+ \frac{1}{2}re^{i\delta })
 & \frac{1}{\sqrt{2}}(1-a)
\end{array}
\right)P\ .~~
\label{eqn:PMNS1}
\end{eqnarray}
where the diagonal matrix $P=\text{Diag}(1,e^{\frac{i\alpha_2}{2}},e^{\frac{i\alpha_3}{2}})$ contains the usual two Majorana phases in the PDG convention\cite{pdg}. The result of Eq. (\ref{eqn:PMNS1}) can be compared to the corrected GR matrix to show how the breaking of the $S$ and $U$ generators corrects the leading order prediction of $r$, $s$, and $a$. However before we make this comparison, phase conventions must be matched. Again, we use the freedom to multiply unphysical phases from the left since they may be absorbed into charged lepton rotations. Multiplying phases from the right corresponds to a redefinition of the Majorana phases. Once this redefinition has been made, the NLO PMNS matrix is in the PDG convention. Then, we may identify the deviation parameters $r$, $a$ and $s$ resulting from the corrected $U_{GR}$ mixing matrix. They are
\begin{align}\label{ranal}
  r &= \sqrt{2}\left|\Delta U_{13}\right|, \\\label{aanal}
  a &= \sqrt{2}\mathrm{Re}\left(\Delta U_{23}\exp\left(\frac{-i \alpha_3}{2}\right)\right), \\\label{sanal}
  s &= \frac{3}{2\phi_g\sqrt{5}}-\frac{1}{2}+3\sqrt{\frac{1}{\phi_g\sqrt{5}}}\mathrm{Re}\left(\Delta U_{12}\exp\left(\frac{-i\alpha_2}{2}\right)\right),
\end{align}
where $\alpha_2$ and $\alpha_3$ are the redefined Majorana phases in the PDG convention\cite{pdg}.  

We pause here to comment on the lack of a concrete prediction even though we have only introduced one parameter.  As remarked earlier the LO neutrino mass matrices given in Eqs. (\ref{eqn:majoranamass}) and (\ref{eqn:diracmass}) respect Form Dominance \cite{FD}.
As such the GR mixing matrix of Eq. (\ref{UGR}) contains no free parameters.  By adding the correction, $\Delta M_{R}$ (cf. Eq. (\ref{eq:majcorr})) to $M_R$, ``one'' additional free parameter is introduced ($y_Cv_4$). However because of the way that the parameter enters, \textit{i.e.} only in the (22) entry of $M_{R}$, the Majorana mass matrix can no longer be diagonalised with a transformation independent of parameters of the model, and the resulting $U_{PMNS}$ will no longer be parameter free.  Hence, at NLO the mixing angles will depend on parameters in the model, as Form Dominance has been broken.

A similar (slightly easier) perturbative exercise may be performed to calculate the NLO corrections to the light neutrino masses.  Performing this calculation reveals the light, complex masses at NLO to be
\begin{eqnarray}
\label{eq:analNLOmass}
\begin{array}{c}
m_1^{NLO}\approx m_1^{LO}-\frac{9 \beta  }{2\phi_g\sqrt{30}   \left(1-3 \phi_g -2 \xi e^{i \delta }  \right)^2}\epsilon, \\\\
m_2^{NLO}\approx m_2^{LO}-\frac{9\beta \phi_g }{2\sqrt{30} \left(2-3 \phi_g +2 \xi e^{i \delta }  \right)^2}\epsilon,\\\\
m_3^{NLO}\approx m_3^{LO}-\frac{9\beta}{2\sqrt{6}(1+4\xi e^{i\delta})^2}\epsilon,
\end{array}
\end{eqnarray}
where the $m_i^{LO}$ are as defined in Eq. (\ref{redmass}).  Notice that the LO sum rule still exists at NLO, \textit{i.e.} 
\begin{equation}
\label{NLOneutmass}
\frac{1}{m_1^{NLO}}+\frac{1}{m_2^{NLO}}=\frac{1}{m_3^{NLO}}.
\end{equation}
This surprising result can be explained by first observing that the NLO correction to the heavy, right-handed neutrino masses will be given by the corresponding diagonal element of the matrix $U_{GR}^T\Delta M_R U_{GR}$.  A calculation of these corrections reveals 
\begin{eqnarray}
\begin{array}{ccc}
M^{NLO}_1=M^{LO}_1+\frac{3 v_4 y_{C}}{\sqrt{5} \phi_g }, & M^{NLO}_2=M^{LO}_2+\frac{3 v_4 \phi_g  y_{C}}{\sqrt{5}},& M^{NLO}_3=M^{LO}_3+3 v_4 y_{C}.
\end{array}
\end{eqnarray}
for the corrected heavy neutrino masses.  Clearly $M^{NLO}_1+M^{NLO}_2=M^{NLO}_3$, preserving the LO sum rule for the heavy right handed neutrino masses of Eq. (\ref{majsumrule}).  Furthermore, because of the form of $m_D$, it can be shown that,
\begin{equation}
m_i=v_u^2 y^2 M_i^{-1}.
\end{equation}
Hence, $M^{NLO}_1+M^{NLO}_2=M^{NLO}_3$ implies $1/m^{NLO}_1+1/m^{NLO}_2=1/m^{NLO}_3$.  In summary, due to the form of the correction to $\Delta M_R$ (cf. Eq. (\ref{eq:majcorr})), the resulting matrix which governs the corrections to the heavy neutrino masses, $U_{GR}^T\Delta M_R U_{GR}$, also shares the same sum rule between its diagonal elements, allowing for the preservation of the sum rule for the heavy neutrino masses at NLO.  Then, because of the form of $m_D$ (cf. Eq. (\ref{eqn:diracmass})), it is possible to deduce that the heavy neutrino masses, $M_i$ are inversely proportional to $m_i$, preserving the light neutrino mass sum rule to $\mathcal{O}(v_4/v_{2,3})$.

Before concluding this section, we briefly discuss NLO corrections to the mass matrices. These will depend on the specific choice of charges $x,y,z$. For the choice $x=-9,y=2,z=5$ we find that, the NLO operators first enter at order $\epsilon^5$, and involve auxiliary flavons, i.e. $\omega$ and $\phi'$, which must be added to aid in the Golden Model's vacuum alignment, cf. Section \ref{sec:alignment}. Examples of such operators are:
\beq
\label{NLOmass}
\frac{1}{\Lambda^5}Lh_uN\lambda\omega^2\varphi^2,~~~~~\frac{1}{\Lambda'\Lambda^4}L h_u Lh_u\lambda \varphi\omega ^2,~~~~\frac{1}{\Lambda^6}Lh_dE \lambda\omega\phi'\varphi^3,~~~~\frac{1}{\Lambda^5}NN\lambda  \varphi^3 \omega^2.
\eeq
where $\Lambda'$ is the Majorana mass scale and $\Lambda$ is the messenger mass scale for the Yukawa sector.  As will be seen in the next section, these additional (auxiliary) flavon fields must be added to aid in the minimisation of the flavon potential.  
Clearly these operators, will contribute to higher order corrections to the neutrino mass matrices and charged lepton mass matrix.
Notice that these results arise when coupling at least more than four flavons to the leptons.  Therefore, the results of the preceding sections are good through order at least $\epsilon^4$.  Furthermore, recall that the charged lepton mass matrix was only approximately diagonal to order $\epsilon^4$, so these additional flavon couplings cause no further problems.


\section{Vacuum Alignment}
\label{sec:alignment}


\begin{table}[h]
	\centering
		\begin{tabular}{|c|c|c||c|c|c|c|c|c|c|}
			\hline
				Field & $\phi'$ & $\omega$ & $\phi^0$  & $\rho^0$ & $\chi^0$ & $\varphi^0$ & $\lambda^0$&$\xi^0$&$\psi^0$ \\ \hline \hline
				$A_5$ & ${\bf 5}$ & ${\bf 4}$ & ${\bf 5}$ & ${\bf 3}$ & ${\bf 4}$ & ${\bf 3'}$ & ${\bf 1}$&${\bf 3'}$&$\bf{5}$ \\ \hline
				$U(1)$ & $-6x$ & $z$ & $6x$ & $11x$ & $8x$ & $2y-z$ & $2y$& $5x+y$&$-2z$  \\ \hline
                                $\!{U}(1)_R\!$ & $0$ & $0$ & $2$ & $2$ & $2$ & $2$ & $2$&$2$&$2$ \\ \hline
		\end{tabular}
	\caption{Auxiliary flavon and driving fields of the $A_5\times U(1)$ model where $x$, $y$, and $z$ are carefully chosen integers (e.g. $x=-9$, $y=2$ and $z=5$). }
\label{tab:driving}
\end{table}

Flavour models of this type must have the alignment of the VEVs of the flavon fields justified by minimising a flavon potential.  Therefore, the explicit VEVs  quoted in Section \ref{sec:model}  must be derived from a flavon potential.  Herein lies the goal of the present section.

To properly align the VEVs of the flavon fields of the $A_5\times U(1)$ Golden Model, a set of auxiliary flavon fields will be added as well as a set of ``driving fields''.  Recall that with the choice of charges $x=-9$, $y=2$ and $z=5$, the auxiliary fields will not contribute to the heavy neutrino mass matrix until at least 6 flavon fields are involved, e.g. with operators of the form $NN\lambda  \varphi^3 \omega^2/\Lambda^5$, cf. Eq.~(\ref{NLOmass}).  They will also not contribute to the Dirac neutrino mass matrix until 5 flavons are involved in operators like $Lh_uN\lambda\omega^2\varphi^2/\Lambda^5$, and they will not couple to the charged lepton mass matrix until at least 6 flavons are involved in operators like $Lh_dE \lambda\omega\phi'\varphi^3/\Lambda^6$.  Notice that there is also a coupling like $L h_u Lh_u\lambda \varphi\omega ^2/\Lambda'\Lambda^4$ contributing to the effective neutrino mass matrix.  As previously discussed, the suppression of these operators provides negligible corrections to the mass matrices.  
Driving fields in turn are similar to flavons in that they are gauge singlets and transform in a nontrivial way under $A_5\times U(1)$.  However, their difference becomes apparent when an additional $U(1)_R$ symmetry is introduced.\footnote{$U(1)_R$ is broken to R-parity when supersymmetry breaking terms are included.}  Under this symmetry, we define all chiral superfields containing Standard Model fermions to have a  $U(1)_R$ charge of $+1$.   Then, all chiral superfields containing Higgs and flavon fields have a $U(1)_R$ charge of $0$.  Driving fields will have a $U(1)_R$ charge of $+2$.  Since the superpotential carries a $+2$ $U(1)_R$ under these conventions, the driving fields may only couple linearly to flavon fields.  
Furthermore, we will work in the supersymmetry preserving limit so that to minimise the flavon potential it is only necessary to enforce that the $F$-terms of the driving fields vanish identically.  These so-called $F$-term conditions will give rise to the vacuum alignments.  The driving fields, additional flavon fields, and their transformation properties under $A_5\times U(1)\times U(1)_R$ can be found in Table \ref{tab:driving}.

To begin the minimisation of the flavon potential, the first step is the construction of the the flavon superpotential from the fields' transformation properties given in Tables \ref{tab:matter} and \ref{tab:driving}.  A straightforward calculation shows the ``LO'' flavon superpotential may be constructed from contractions of the operators

\begin{equation}
\label{nocont}
\phi'\phi^0,~\phi\phi\phi^0,~\theta\chi\phi^0,~\phi'\chi\rho^0,\phi\chi\chi^0,~\omega\omega\psi^0~,~\omega\varphi\varphi_0~,~~\lambda\lambda\lambda^0,~\text{and}~\lambda\chi\xi^0.
\end{equation}
Because some of these operators have multiple independent contractions that can result from them, it is necessary to explicitly write the flavon superpotential in terms of the relevant contractions.  Doing this yields

\begin{equation}
\begin{split}
W_d&=M\left(\phi'\phi^0\right)_{\bf 1_s}+g_1\left( \left(\phi\phi\right)_{\bf 5_{1,s}}\phi^0\right)_{\bf 1_s}+g_2\left( \left(\phi\phi\right)_{\bf 5_{2,s}}\phi^0\right)_{\bf 1_s}+g_3\left(\theta\left( \chi\phi^0\right)_{\bf 1_s}\right)_{\bf{1_s}} \\
&+g_4\left(\left(\phi'\chi\right)_{\bf 3_a}\rho^0\right)_{\bf 1_s}+g_5\left(\left(\phi\chi\right)_{\bf 4_s}\chi^0\right)_{\bf 1_s}+g_6\left(\left(\phi\chi\right)_{\bf 4_a}\chi^0\right)_{\bf 1_s}+g_7\left(\left(\omega\omega \right)_{\bf{5_s}}\psi^0\right)_{\bf{1_s}}\\
&+g_8\left(\left(\omega\varphi\right)_{\bf 3'}\varphi_0\right)_{\bf 1_s} + +g_{9}\left(\left(\lambda\lambda\right)_{\bf 1_s}\lambda^0\right)_{\bf 1_s}+g_{10}\left(\left(\lambda\chi\right)_{\bf{3'}}\xi^0\right)_{\bf 1_s}.
\label{eqn:driving}
\end{split}
\end{equation}
As previously discussed, the superpotential can be used to construct the driving fields' $F$-terms, that when vanish, minimise the flavon potential.  The vanishing $F$-term of $\rho^0$ and $\chi^0$ that provide the alignment for the VEVs of $\phi$, $\phi'$, and $\chi$ are 
\begin{eqnarray}
\begin{split}
\frac{\partial W_d}{\partial \rho^0_1}&=g_4\left(-\phi'_5\chi_2-2\phi'_4\chi_3+2\phi'_3\chi_4+\phi'_2\chi_5\right)=0 \\
\frac{\partial W_d}{\partial \rho^0_2}&=g_4\left(-\sqrt{3}\phi'_5\chi_1-\sqrt{2}\phi'_4\chi_2+\sqrt{2}\phi'_2\chi_4+\sqrt{3}\phi'_1\chi_5\right)=0 \\
\frac{\partial W_d}{\partial \rho^0_3}&=g_4\left(\sqrt{3}\phi'_2\chi_1-\sqrt{3}\phi'_1\chi_2-\sqrt{2}\phi'_5\chi_3+\sqrt{2}\phi'_3\chi_5\right)=0 \\
\frac{\partial W_d}{\partial \chi^0_1}&=g_5 \left(3 \sqrt{2} \chi _1 \phi _5-\sqrt{3} \chi _2 \phi _4+4 \sqrt{3}\chi _3 \phi _3-\sqrt{3} \chi _4 \phi _2+3 \sqrt{2} \chi _5 \phi _1\right)\\
&+g_6 \left(-\sqrt{2} \chi _1 \phi _5+\sqrt{3} \chi _2 \phi _4-\sqrt{3} \chi _4 \phi _2+\sqrt{2} \chi _5 \phi _1\right)=0\\
\frac{\partial W_d}{\partial \chi^0_2}&=g_5 \left(3 \sqrt{2} \chi _1 \phi _4-\sqrt{3} \chi _2 \phi _3-\sqrt{3}\chi _3 \phi _2+3 \sqrt{2} \chi _4 \phi _1+4 \sqrt{3} \chi _5 \phi _5\right)\\
&+g_6 \left(\sqrt{2} \chi _1 \phi _4+\sqrt{3} \chi _2 \phi _3-\sqrt{3} \chi _3 \phi _2-\sqrt{2} \chi _4 \phi _1\right)=0\\
\frac{\partial W_d}{\partial \chi^0_3}&=g_5 \left(3 \sqrt{2} \chi _1 \phi _3+4 \sqrt{3} \chi _2 \phi _2+3\sqrt{2} \chi _3 \phi _1-\sqrt{3} \left(\chi _4 \phi _5+\chi _5 \phi _4\right)\right)\\
&+g_6 \left(\sqrt{2} \chi _1 \phi _3-\sqrt{2} \chi _3 \phi _1+\sqrt{3} \left(\chi _5 \phi _4-\chi _4 \phi _5\right)\right)=0\\
\frac{\partial W_d}{\partial \chi^0_4}&=g_5 \left(3 \sqrt{2} \chi _1 \phi _2+3 \sqrt{2} \chi _2 \phi_1-\sqrt{3} \left(\chi _3 \phi _5-4 \chi _4 \phi _4+\chi _5 \phi _3\right)\right)\\
&+g_6 \left(-\sqrt{2} \chi _1 \phi _2+\sqrt{2} \chi _2 \phi _1+\sqrt{3} \left(\chi _5 \phi _3-\chi _3 \phi _5\right)\right)=0
\end{split}
\label{eqn:coupledeqs}
\end{eqnarray}
which have the non-vanishing solutions necessary for spontaneous symmetry breaking:
\begin{equation}
 \vev{\chi}=\left(0,0,v_{\chi},0,0\right)^T, \; \vev{\phi}=\left(0,0,0,v_{\phi},0\right)^T \; \mathrm{and} \; \vev{\phi'}=\left(0,v_{\phi'},v_{\phi'},0,0\right)^T.
\label{eqn:clepvevs}
\end{equation}

 We turn now to the $F$-term conditions resulting from $\phi^0$.  These are found to be
\begin{equation}
\begin{split}
\frac{\partial W_d}{\partial \phi^0_1}&=M\phi'_1+2g_1\left(\phi_1^2+\phi_2\phi_5-2\phi_3\phi_4\right)+2g_2\left(\phi_1^2-2\phi_2\phi_5+\phi_3\phi_4\right)+g_3\theta\chi_1=0, \\
\frac{\partial W_d}{\partial \phi^0_2}&=M\phi'_5+2g_1\left(\phi_1\phi_5+\sqrt{6}\phi_2\phi_4\right)+g_2\left(-4\phi_1\phi_5+\sqrt{6}\phi_3^2\right)+g_3\theta\chi_5=0, \\
\frac{\partial W_d}{\partial \phi^0_3}&=M\phi'_4+g_1\left(-4\phi_1\phi_4+\sqrt{6}\phi_5^2\right)+2g_2\left(\phi_1\phi_4+\sqrt{6}\phi_2\phi_3\right)+g_3\theta\chi_4=0, \\
\frac{\partial W_d}{\partial \phi^0_4}&=M\phi'_3+g_1\left(-4\phi_1\phi_3+\sqrt{6}\phi_2^2\right)+2g_2\left(\phi_1\phi_3+\sqrt{6}\phi_4\phi_5\right)+g_3\theta\chi_3=0, \\
\frac{\partial W_d}{\partial \phi^0_5}&=M\phi'_2+2g_1\left(\phi_1\phi_2+\sqrt{6}\phi_3\phi_5\right)+g_2\left(-4\phi_1\phi_2+\sqrt{6}\phi_4^2\right)+g_3\theta\chi_2=0.
\end{split}
\label{eqn:coupledeqs2}
\end{equation}
Applying the previous results for the alignments of $\vev{\chi}$, $\vev{\phi}$, and $\vev{\phi'}$, collapses these equations to the rather simple constraints
\begin{equation}
\label{simpleF}
M v_{\phi'}+g_3v_{\theta}v_{\chi}=0~\text{and}~M v_{\phi'}+\sqrt{6}g_2v_{\phi}^2=0
\end{equation}
which together imply $v_{\phi}^2=\frac{g_3}{\sqrt{6}g_2}v_{\theta}v_{\chi}$.  Notice that since $v_{\chi}\neq 0$ and $v_{\phi}\neq 0$, then $v_{\theta}\neq 0$.
  
To find the other nontrivial solutions of Eqs.~(\ref{eqn:coupledeqs}) and (\ref{eqn:coupledeqs2}) which yield the same relationship between the VEVs of $\phi$, $\theta$ and $\chi$, i.e. $v_{\phi}^2=\frac{g_3}{\sqrt{6}g_2}v_{\theta}v_{\chi}$, we must only act with all $A_5$ group elements on the VEVs given in 
Eq.~(\ref{eqn:clepvevs}) and verify which still solve the F-term conditions given in Eqs.~(\ref{eqn:coupledeqs}) and (\ref{eqn:coupledeqs2}), while keeping the desired aforementioned relationship between the flavon VEVs.  It has been verified that 17 other nontrivial $A_5$ group elements acting on the preceding VEV alignments produce $v_{\phi}^2=\frac{g_3}{\sqrt{6}g_2}v_{\theta}v_{\chi}$, providing a complete set of degenerate, non-vanishing solutions (for this particular choice of alignment).\footnote{These solutions can be obtained by the action of $T$, $T^2$, $T^3$, $T^4$, $S$, $U$, $ST$, $TS$, $UT$, $TU$, $SU$, $T^2S$, $T^2U$, $UT^2$, $T^3S$, $T^3ST^3$, or $T^3ST^4$ on the VEV alignments in Eq. (\ref{eqn:clepvevs}).} The other $60-(17+1)=42$ elements acting on the VEV alignments in Eq. (\ref{eqn:clepvevs}) fail to produce the same relationship between the flavon VEVs.
Thus, they will be neglected.

With the justification of the alignment of $\vev{\chi}$, $\vev{\phi}$, and $\vev{\theta}$, we turn ourselves to aligning the flavons associated with the neutrino sector.

Aligning the flavons associated with the neutrino sector we begin by minimising the $F$-terms associated with $\psi^0$.  We find the following conditions on the alignment of the auxiliary field $\omega$:
\begin{equation}
\begin{split}
&\frac{\partial W_d}{\partial \psi^0_1}=g_7\left( 2 \sqrt{3} \omega _1 \omega _4-2 \sqrt{3} \omega _2 \omega _3\right)=0,\\
&\frac{\partial W_d}{\partial \psi^0_2}=g_7\left(2 \sqrt{2} \omega _2^2-2 \sqrt{2} \omega _1 \omega _3\right)=0,\\
&\frac{\partial W_d}{\partial \psi^0_3}=g_7\left(2 \sqrt{2} \omega _1 \omega _2-2 \sqrt{2} \omega _4^2\right)=0,\\
&\frac{\partial W_d}{\partial \psi^0_4}=g_7\left(2 \sqrt{2} \omega _3 \omega _4-2 \sqrt{2} \omega _1^2\right)=0,\\
&\frac{\partial W_d}{\partial \psi^0_5}=g_7\left(2 \sqrt{2} \omega _3^2-2 \sqrt{2} \omega _2 \omega _4\right)=0.\\
\label{eqn:omegaalign}
\end{split}
\end{equation}
which have nontrivial solutions of the form
\begin{eqnarray}
\langle\omega\rangle\propto \rho^m T_4^n\left(
\begin{array}{c}
1\\
1\\
1\\
1
\end{array}\right)=
\rho^m\left(
\begin{array}{cccc}
 \rho  & 0 & 0 & 0 \\
 0 & \rho ^2 & 0 & 0 \\
 0 & 0 & \rho ^3 & 0 \\
 0 & 0 & 0 & \rho ^4
\end{array}
\right)^n\left(
\begin{array}{c}
1\\
1\\
1\\
1
\end{array}\right)
\end{eqnarray}
in which $\rho=e^{\frac{2\pi i}{5}}$ and $m,n=0,1,2,3,4$.  For the rest of this work, we take $m=n=0$ such that $\langle\omega\rangle=v_{\omega}(1,1,1,1)^T$.

With the alignment for $\langle\omega\rangle$ calculated, we turn now to the $F$-terms associated with $\varphi^0$ and find the following $F$-term conditions: 
\begin{equation}
\begin{split}
&\frac{\partial W_d}{\partial \varphi^0_1}=g_8 \left( \sqrt{2} \varphi _5 \omega _1+2\sqrt{2} \varphi _4 \omega _2-2 \sqrt{2} \varphi _3 \omega _3- \sqrt{2} \varphi _2 \omega _4\right)=0, \\
&\frac{\partial W_d}{\partial \varphi^0_2}=g_8 \left(-2 \varphi _3 \omega _1+\varphi _2 \omega _2+\sqrt{6} \varphi _1 \omega _3-3\varphi _5 \omega _4\right)=0, \\
&\frac{\partial W_d}{\partial \varphi^0_3}=g_8 \left(3\varphi _2 \omega _1-\sqrt{6} \varphi _1 \omega _2-\varphi _5 \omega _3+2 \varphi _4 \omega _4\right)=0,
\label{eqn:nualign1}
\end{split}
\end{equation}
Since $v_{\omega}\neq 0$ and by assuming the previous solution for $\langle\omega\rangle$, the above equations yield
the solution
\begin{equation}
\vev{\varphi}=\left(\sqrt{\frac{2}{3}}\left(v_2+v_3\right),v_2,v_3,v_3,v_2\right)^T. 
\end{equation}
The remaining set of $F$-terms to analyse belong to $\lambda^0$ and $\xi^0$.  They are
\begin{equation}
\begin{split}
&\frac{\partial W_d}{\partial\lambda^0}=g_{9} \left(\lambda_1^2+2 \lambda_2 \lambda_3\right),\\
&\frac{\partial W_d}{\partial\xi^0_1}= g_{10} \left(\sqrt{3} \lambda _1 \chi _1+\lambda _3 \chi _2+\lambda _2 \chi _5\right) \\
&\frac{\partial W_d}{\partial\xi^0_2}=  g_{10} \left(-\sqrt{2} \lambda _2 \chi _3+\lambda _1 \chi _4-\sqrt{2} \lambda _3 \chi _5\right)  \\
&\frac{\partial W_d}{\partial\xi^0_3}=  g_{10} \left(-\sqrt{2} \lambda _2 \chi _2+\lambda _1 \chi _3-\sqrt{2} \lambda _3 \chi _4\right) 
\end{split}
\end{equation}
Upon applying the previously found alignment for $\langle\chi\rangle$ (\textit{i.e.} $\langle\chi\rangle\propto (0,0,1,0,0)^T$), these $F$-terms vanish when $\langle\lambda\rangle$ is aligned as
\begin{eqnarray}
\langle\lambda\rangle\propto
\left(\begin{array}{c}
0\\
0\\
1
\end{array}\right),
\end{eqnarray}
the $Z_2^S\times Z_2^U$ breaking alignment for $\langle\lambda\rangle$ used in Section \ref{sec:theta13}.  Before concluding the discussion of the flavon potential, we pause to comment on the generality of the $A_5\times U(1)$ model's vacuum solutions.


Because the alignment of $\vev{\omega}$ was not dependent on the results of the alignments for $\vev{\chi}$, $\vev{\phi}$, and $\vev{\phi'}$ it is possible to begin the consideration of possible $A_5$ vacua by considering the alignment of $\vev{\omega}$, which may be brought to the desired $(1,1,1,1)^T$ by action of the $T$ generator.  Then, having properly aligned $\vev{\omega}$, the VEV alignment of $\vev{\varphi}$ is determined uniquely.  Then, the VEVs of the flavons associated with the charged leptons (\textit{i.e.} $\chi$, $\phi$, and $\phi'$) may be brought to the desired form by application of relevant powers of $S$, $T$, and $U$ matrices.  In general, these transformations will change the already aligned $\vev{\omega}$ by multiples of $T$.  This in turn will force $\vev{\varphi}$ out of the alignment needed for GR mixing.  Therefore, Nature is required to choose one of five vacua necessary for the Golden Model.  This is a mild assumption.  It is worth pointing out that we have not justified the relative size of each of the flavon VEVs with respect to each other. Hence, in this paper we impose by hand the relative sizes of the VEVs with respect to each other.  Said in a different way, there are 7 flavon field VEVs (8 flavon VEV parameters due to the presence of $v_2$ and $v_3$ in $\langle\varphi\rangle$) and one equation relating 3 of the VEVs, i.e. $v_{\phi}^2=\frac{g_3}{\sqrt{6}g_2}v_{\theta}v_{\chi}$.  Therefore, this model has at least 7 complex flat directions corresponding to the undetermined complex VEVs. 

Before discussing the phenomenological predictions of the Golden Model, we pause here to discuss the NLO corrections to the flavon superpotential.  The NLO contributions coming from the charges in the Tables \ref{tab:matter} and \ref{tab:driving} with  $x=-9$, $y=2$ and $z=5$ enter in when 3 flavons and 1 driving field are involved, for example
\beq
\frac{1}{\Lambda}\rho^0\chi\phi^2,  \ \ \ \ \frac{1}{\Lambda} \psi^0\theta\omega \varphi.
\eeq
Notice that the first operator in the above equation exists for any integer value for $x$, $y$, and $z$, but the latter does not.  Further observe that in particular there are no additional renormalisable operators which are introduced. Thus, we shall assume that the above
non-renormalisable operators give a negligible contribution to the vacuum alignment, for example due to a heavier messenger mass scale here than in the Yukawa sector.


\section{Phenomenological Predictions}
\label{sec:sumruleimp}


In this section of the paper, we study the phenomenological implications of the complex mass sum rule of Eqs. (\ref{sumrule}) and (\ref{NLOneutmass}). Although this sum rule has been studied before in $S_4$ and $A_4$ flavour symmetry based models\cite{sumruleorig}, this work was done in the era of a small reactor angle.  There has been some mass sum rule studies after the advent the measurement of a large value for $\theta_{13}$\cite{sumruleafter}, but this work did not go into great detail for the inverse mass sum rule we have generated in our model.  Hence, we are motivated to perform our own analysis.

\begin{figure}[ht!]
\centering
\includegraphics{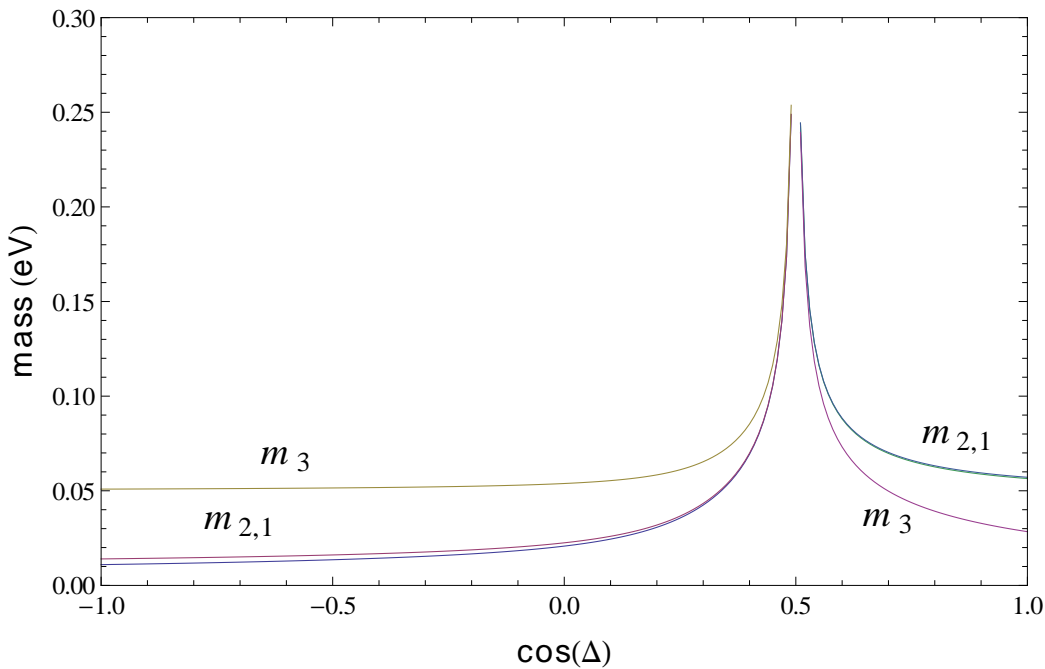}
\caption{mass vs. $\text{cos(}\Delta\text{)}$ for the normal and inverted neutrino mass orderings}
\label{fig:massplot}
\end{figure}

We begin the analysis of the phenomenological implications of the Golden Model by re-expressing the complex mass sum rules of Eqs. (\ref{sumrule}) and (\ref{NLOneutmass}) in terms of the physical neutrino masses, $|m_i|$, and Majorana phases, $\alpha_2$ and $\alpha_3$ (note the sign convention change on $\alpha_i$):
\begin{equation}
\label{sumrulewphase}
\frac{1}{|m_1|}+\frac{e^{-i\alpha_2}}{|m_2|}=\frac{e^{-i\alpha_3}}{|m_3|}
\end{equation}
where the superscripts ``LO'' and ``NLO'' have been dropped on $m_i$.  From the above formula, it is easy to see that Eq. (\ref{sumrulewphase}) implies that
\begin{equation}
\label{sumrulecos}
\frac{1}{|m_1|^2}=\frac{1}{|m_2|^2}+\frac{1}{|m_3|^2}-\frac{2\cos{\Delta}}{|m_2||m_3|},
\end{equation}
where $\Delta=\alpha_2-\alpha_3$.  With the realisation of this form, it becomes clear that together with the two experimentally measured values of the neutrino mass squared differences $\Delta m_{21}^2$ and $|\Delta m_{23}^2|$, where $\Delta m_{ij}^2=|m_i|^2-|m_j|^2$, it is possible to numerically calculate the individual neutrino masses for a specified value of $\Delta$.  The results of this analysis can be found in Figure \ref{fig:massplot}.  The plot of Figure \ref{fig:massplot} was made using the central value mass squared differences of Ref. \cite{global}.  Using another global fit or including a $3\sigma$ error band on the masses would produce negligible plot differences or slightly thicker lines for the light neutrino masses, respectively. The former results from the relatively small variance between the global fits' values for the $\Delta m_{ij}^2$, and the latter results from the $3\sigma$ deviation being not large enough to produce a noticeable difference on the plot when considering the inverse mass sum rule, i.e. the same behaviour.

The plot of Figure \ref{fig:massplot} contains many interesting features.  The first of which is the discontinuity at $\cos(\Delta)\approx .5$.  Namely, given the experimentally measured values for $\Delta m_{ij}^2$ and the neutrino sum rule from Eq. (\ref{sumrulecos}) no solution exists. The second interesting feature is that only normal ordered neutrino masses can only exist below $\cos(\Delta)\approx .5$, and inverted only above $\cos(\Delta)\approx .5$.  This implies that varying $\Delta$ provides a transition between normal and inverted neutrino mass orderings, the justification of why there is no solution for neutrino masses at $\cos(\Delta)\approx .5$.

The physical masses as a function of $\cos(\Delta)$ serve as a stepping stone for numerically calculating the values of $\xi$, $\delta$, and $|\beta|$ (cf. Eq. (\ref{redmass})) for a fixed value of $\cos(\Delta)$.  To correctly calculate these parameters to NLO, it is necessary to use the NLO masses of Eq. (\ref{eq:analNLOmass}).  However, notice that these masses include an additional complex parameter $\epsilon=|\epsilon|e^{i\theta_{\epsilon}}$.  We find $|\epsilon|$ by demanding that the TB deviation parameter $r$ (cf. Eq. (\ref{ranal})) matches the corresponding value derived from the central value of the reactor angle given in Ref. \cite{global}, \textit{i.e.} $r=.22$.  Then, $|\epsilon|$ may be solved for as a function of the $\xi$, $\delta$, $|\beta|$ parameters and removed from the NLO masses of Eq. (\ref{eq:analNLOmass}) leaving 4 unknown real parameters.  $\theta_{\epsilon}$ can be eliminated by minimising and maximising the masses with respect to $\theta_{\epsilon}$.  After this, the maximised and minimised NLO masses may be expressed in terms of only $\xi$, $\delta$, and $|\beta|$.  Setting these analytic forms for the maximised and minimised NLO neutrino masses equal to their corresponding values found in Figure \ref{fig:massplot} reveals the values of the parameters $\xi$, $\delta$, and $|\beta|$ as a function of $\cos(\Delta)$.\footnote{Numerical instabilities in some of the solutions for the $\xi$, $\delta$, and $|\beta|$ results of the minimised NLO neutrino masses urged us to use the maximised results in our analysis.  Although, the stable results show little difference between the two cases.} Notice that the parameters entering in the masses also appear in the analytic forms of the corrected mixing matrix elements, cf. Eqs. (\ref{deltaUbegin})-(\ref{correctionend}).  Therefore, it is possible to numerically calculate the elements of the PMNS matrix and find the TB deviation parameters $s$ and $a$ of Eqs. (\ref{aanal})-(\ref{sanal}) as a function of $\cos(\Delta)$.

\subsection{NLO Corrections to $s$ and $a$}
\begin{figure}[ht!]
\centering
\includegraphics{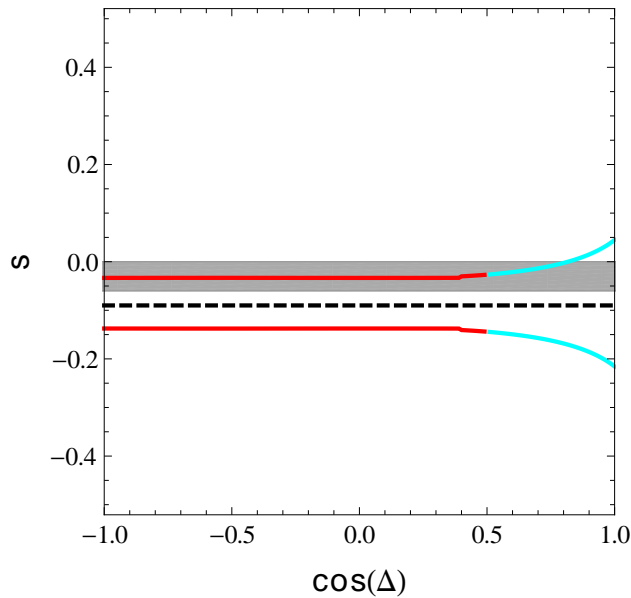}
\caption{s vs. $\text{cos(}\Delta\text{)}$ for fixed $r=.22$:  The region between the red lines represents the possible values the $s$ parameter can take for normal ordered neutrino masses, and the region between the light blue lines represents the possible values for the inverted ordering. The thickness of the lines takes into account the $1\sigma$ ranges for the $\Delta m_{ij}^2$ used in producing the plot, and the grey area represents the experimentally allowed $1\sigma$ range from Eq. (\ref{rsafit}).  The LO Golden Ratio Prediction is given by the dashed line.}
\label{splot}
\end{figure}

\begin{figure}[ht!]
\centering
\includegraphics{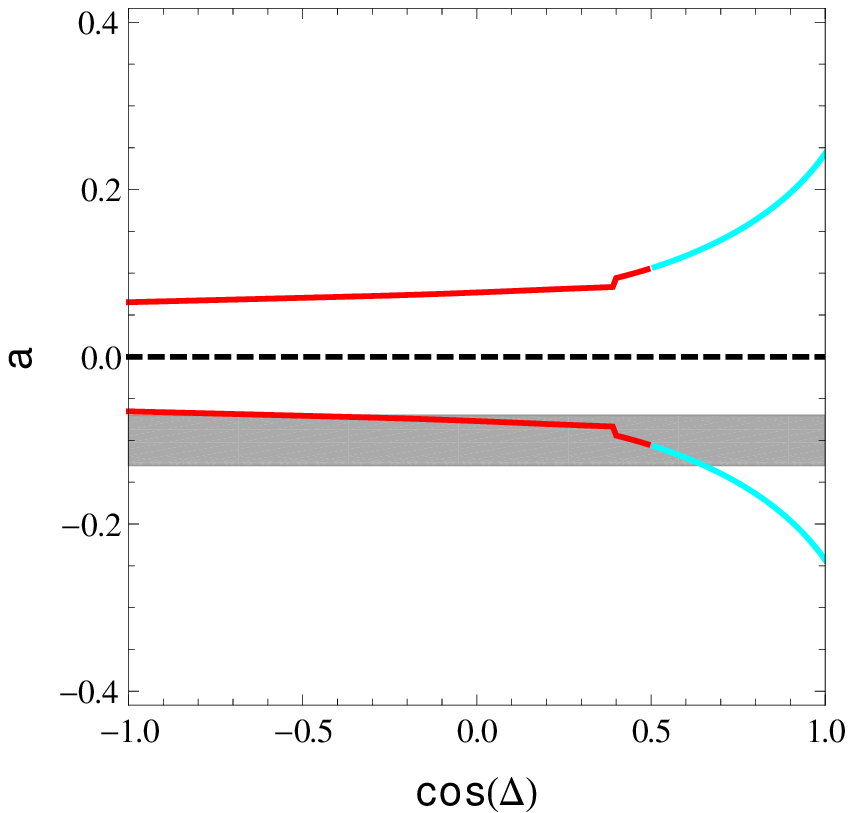}
\caption{a vs. $\text{cos(}\Delta\text{)}$ for fixed $r=.22$:  The region between the red lines represents the possible values the ``a" parameter can take for the normal ordered neutrino masses, and the region between the light blue lines represents the possible values for the inverted ordering. The thickness of the lines takes into account the $1\sigma$ ranges for the $\Delta m_{ij}^2$ used in producing the plot, and the grey area represents the experimentally allowed $1\sigma$ range from Eq. (\ref{rsafit}).  The LO Golden Ratio Prediction is given by the dashed line.}
\label{aplot}
\end{figure}

In this section, we report the results obtained from using the previously discussed method for determining $|\epsilon|$, $|\beta|$, $\xi$, and $\delta$ to calculate the TB deviation parameters $s$ and $a$ to NLO.  Notice that in the definitions of $a$ and $s$ in Eqs. (\ref{aanal})-(\ref{sanal}), the unknown Majorana phases $\alpha_2$ and $\alpha_3$ appear.  $s$ and $a$ are minimised and maximised with respect to these phases to remove their dependence on them.  Then, the calculation of how the $s$ and $a$ parameters vary as a function of $\cos(\Delta)$ is straightforward.  The plots resulting from parameter scans ranging over values of $\cos(\Delta)\in [-1,1]$ for the $s$ and $a$ parameters can be found in Figures \ref{splot} and \ref{aplot}, respectively.

The plots in Figures \ref{splot} and \ref{aplot} show some interesting features.  The first feature to notice is that the LO GR prediction is contained in the allowed region predicted by the NLO Golden Model.  This is due to the fact that if $\Delta U_{12} e^{-i\frac{\alpha_2}{2}}$ and $\Delta U_{23} e^{-i\frac{\alpha_3}{2}}$ are purely imaginary then the corrections to $s$ and $a$ vanish, respectively, as can be seen from Eqs. (\ref{aanal})-(\ref{sanal}).  The second slightly more interesting feature to notice is that the at $\cos(\Delta)=.5$ there exists no solution, a relic of the mass vs. $\cos(\Delta)$ analysis.\footnote{It should be noted that in the graph of $a$ vs. $\cos(\Delta)$, the value of $a$ in the inverted hierarchy takes larger values which begin to break down the $r$, $s$, $a$, perturbative expansion for the PMNS matrix in $a$.}  Yet, the most intriguing result of this analysis is that the values for $s$ and $a$ can actually lie within the $1\sigma$ band of Eq. (\ref{rsafit}).  Clearly, the NLO corrections provide the needed adjustment to the Golden Model's problematic LO predictions of $r=a=0$ and $s=-.09$, by allowing them to be shifted to agree with current experimental bounds.

\begin{figure}[ht!]
\centering
\includegraphics{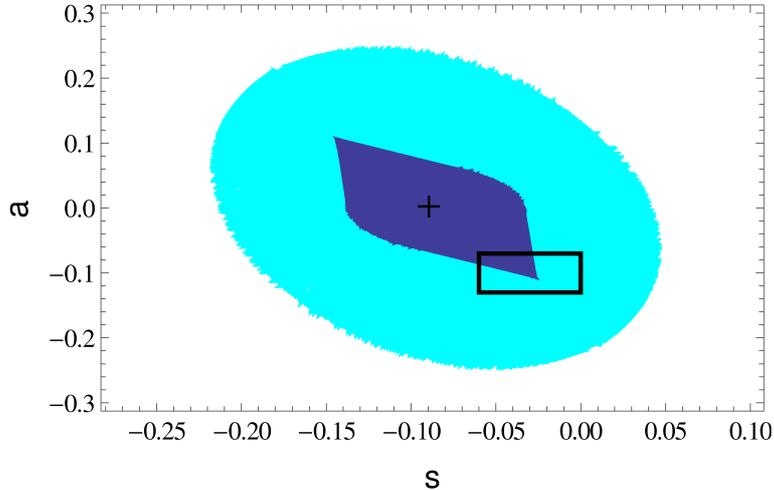}
\caption{a vs. s for fixed $r=.22$: Purple represents predictions from normally ordered neutrino masses and the light blue points from inverted neutrino masses.  The purple region is also where they overlap.  The black rectangle designates the allowed $1\sigma$ region of Eq. (\ref{rsafit}) and the black cross is the LO Golden Ratio prediction. }
\label{avsinvnorm}
\end{figure}

Having calculated the allowed ranges which $s$ and $a$ can take (as a function of $\cos(\Delta)$), there still exists another plot which needs to be generated to further probe the predictions of the Golden Model, $a$ vs. $s$. The plot of $a$ vs. $s$ will further restrict our parameter space of the Golden Model through the correlation of $a$ and $s$.  The method for generating this plot is straightforward because all relevant parameters are known with the exception of the Majorana phases, $\alpha_2$ and $\alpha_3$, and the phase of $\epsilon$, $\theta_{\epsilon}$.  Unfortunately $\Delta=\alpha_2-\alpha_3$ does not enter in either of the forms of the $a$ or $s$ parameters in Eqs. (\ref{aanal})-(\ref{sanal}), so the method of taking the maximum and minimum values cannot be used here because we want a value of $a$ for each $s$.  Thus, it is necessary to introduce another parameter $\Sigma$=$\alpha_2+\alpha_3$, so that the parameter $\Delta$ may be utilised when analysing how $s$ and $a$ vary with each other.  Doing this enables $\alpha_2$ and $\alpha_3$ to be re-expressed as $\alpha_2=1/2(\Delta+\Sigma)$ and $\alpha_3=1/2(\Sigma-\Delta)$. Since all parameters are known for each value of $\Delta$, all that is left unknown is $\Sigma$ and $\theta_{\epsilon}$. Generating random values for $\Sigma$,$\theta_{\epsilon}\in [0, 2\pi]$, while keeping the consistency of the definition of the Majorana phases, yields the plot found in Figure \ref{avsinvnorm}.  As can be seen, this plot further constrains the parameter space of the Golden Model, as it contains a subset of the values displayed in the plots of Figures \ref{splot} and \ref{aplot}, through the correlation of $a$ and $s$.  Yet even though the correlation of $a$ and $s$ further reduces the Golden Model's parameter space, the experimentally determined values for $a$ and $s$ can be still accommodated by the Golden Model, providing consistent phenomenological predictions which are able to match the bounds of Eq. (\ref{rsafit}).  However, to further analyse the phenomenology of this model, we next consider the Golden Model's predictions for neutrinoless double beta decay.

\subsection{Neutrinoless Double Beta Decay}
\begin{figure}[ht!]
\centering
\includegraphics{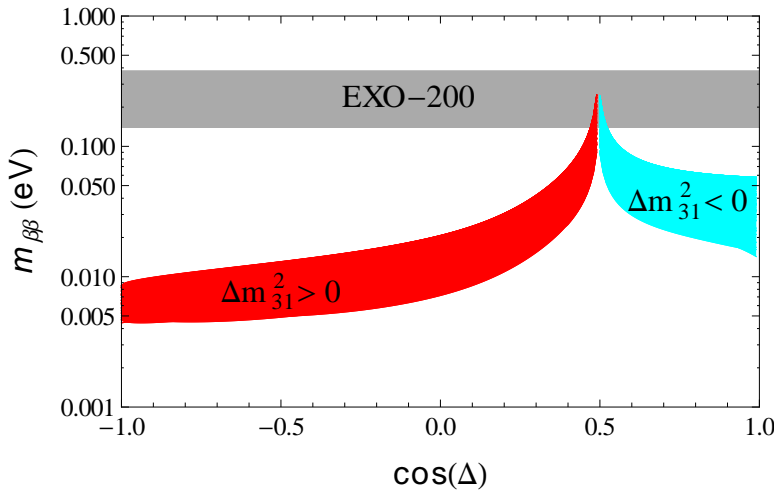}
\caption{$m_{\beta\beta}$ vs. $\text{cos(}\Delta\text{)}$: The red region corresponds to the values that $m_{\beta\beta}$ can take in the normal neutrino mass ordering, the light blue region corresponds in the inverted ordering, and the grey region is the experimental range of upper bounds that $m_{\beta\beta}$ can take from EXO-200\cite{mbb}.}
\label{doublebeta}
\end{figure}

From the plots in Figures \ref{splot} and \ref{aplot}, it is seen that as the parameter $\Delta$ varies, a transition between normal ordered and inverted ordered neutrino masses occurs.  Therefore, it is interesting to analyse the behaviour of the effective Majorana mass parameter, $m_{\beta\beta}$, as a function of $\cos(\Delta)$, where
\begin{equation}
\label{effmajmass}
m_{\beta\beta}=| |m_1|U_{11}^2+|m_2|U_{12}^2e^{-i\alpha_2}+|m_3|U_{13}^2e^{-i\alpha_3}|.
\end{equation}
Notice that in Eq. (\ref{effmajmass}), $U_{ij}$ represent the entries of the PMNS matrix without Majorana phases.  This analysis is preformed by utilising the previously discussed parameters $\Delta=\alpha_2-\alpha_3$ and $\Sigma=\alpha_2+\alpha_3$, so that Eq. (\ref{effmajmass}) may be re-expressed in terms of $\Sigma$ and $\Delta$, instead of $\alpha_2$ and $\alpha_3$.  Then, it is possible to invoke the work from previous sections to numerically generate a plot revealing how $m_{\beta\beta}$ changes as $\cos(\Delta)$ varies.  The results from this analysis can be found in Figure \ref{doublebeta}.  Notice that (as postulated), $\cos(\Delta)$ ``disentangles" degenerate neutrino orderings, as can be seen from the transition at $\cos(\Delta)=.5$.  Also, the results are consistent with the  experimental upper bounds on $m_{\beta\beta}$ reported by EXO-200\cite{mbb}.  Yet the most important conclusion one can draw from Figure \ref{doublebeta} is that the Golden Model is testable in the near future.  Due to the neutrino mass sum rule of Eqs. (\ref{sumrule}) and (\ref{NLOneutmass}), the normal neutrino mass ordering's prediction for $m_{\beta\beta}$ cannot be arbitrarily small.  Hence, the model has the capability of being tested by the next generation of neutrinoless double beta decay experiments.

\begin{figure}[ht!]
\centering
\includegraphics{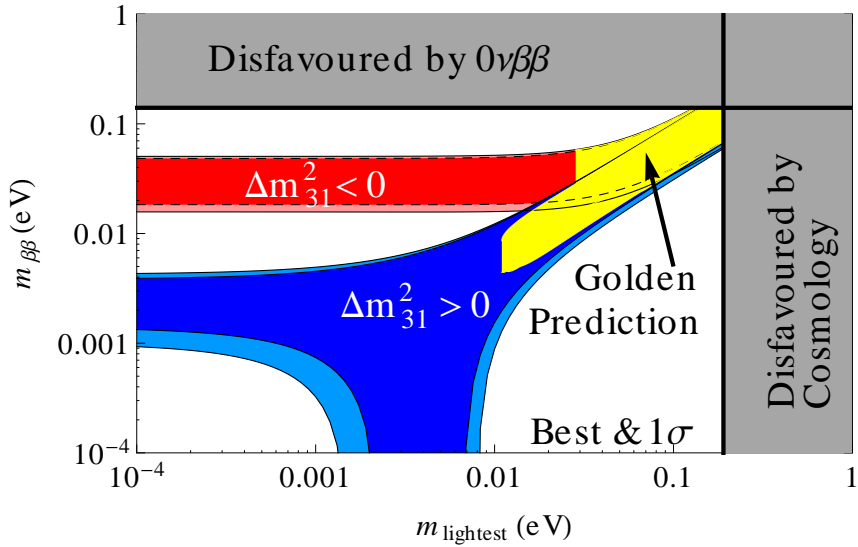}
\caption{$m_{\beta\beta}$ vs. $m_{lightest}$:  The red and light red regions represent the model independent values that the inverted neutrino mass ordering can take based on the central value and 1$\sigma$ deviation of Ref. \cite{global}, respectively.  The blue and light blue regions are the analogue of this for the normal neutrino mass ordering. The gold regions correspond to the Golden Model's prediction for $m_{\beta\beta}$ in both the normal and inverted orderings.}
\label{doublebeta2}
\end{figure}

We continue the analysis of the Golden Model's phenomenology by constructing, $m_{\beta\beta}$ vs. $m_{lightest}$.  This is possible to do since the neutrino masses and $m_{\beta\beta}$ are known for each value of $\text{cos(}\Delta\text{)}$. Thus, it is straightforward to find the maximum and minimum values of $m_{\beta\beta}$ as well as the value of the lightest neutrino mass (\textit{i.e.} $m_1$ in the normal ordering and $m_3$ in the inverted ordering) for each value of $\text{cos(}\Delta\text{)}$.  These results can then be combined to produce the gold regions of the plot found in Figure \ref{doublebeta2}.  The plot in Figure \ref{doublebeta2} also contains blue and red regions corresponding to the normal and inverted neutrino mass orderings if there was no neutrino mass sum rule.  The different shades of the blue and red regions correspond to the the different ranges obtained when taking into account the best fit and  $1\sigma$ deviations away from the best fit.\footnote{It is important to note that in order to generate the model independent case of Figure \ref{doublebeta2}, we have used the code developed in Ref. \cite{merle} with the global fit values of Ref. \cite{global}.}  It should be clear from Figure \ref{doublebeta2} that the Golden Model's inverse mass sum rule severely restricts the allowed values that $m_{\beta\beta}$ can take,
leading to a lower bound of $m_{\rm lightest}\simgt 0.01$ eV and $m_{\beta\beta}\simgt 0.005$ eV
for the normal mass ordering, with the mass spectrum extending into the quasi-degenerate region.  As previously discussed, this allows the Golden Model to be tested by near future neutrinoless double beta decay experiments. 

Before concluding, we comment on the absence of the Dirac CP violating phase in the preceding discussion.  Since we are working within the standard PDG parametrisation of the PMNS matrix, it can be seen that the last 
term in Eq. (\ref{effmajmass}) involves both the Dirac phase and a Majorana phase. In the plots we have 
considered this term to be described by only one independent phase. 
Hence, the neutrinoless double beta decay plots do not depend on the value 
of the Dirac CP phase in our approach.  However we also briefly mention that the Golden Model is consistent with a Jarlskog Invariant of 
$J=\text{Im}(U_{11}U_{33}U_{13}^*U_{31}^*)\approx .05 \sin (\delta)$.

\newpage


\section{Conclusions}
\label{sec:conclusion}


The Golden Ratio mixing prediction of a zero reactor angle means that this model, along with other simple schemes such as TB mixing, is no longer viable. Similarly, there is increasing evidence that the atmospheric angle lies in the first octant, and so the prediction of maximal atmospheric mixing, also shared by TB mixing, is also disfavoured.
However, while the TB prediction for the solar angle remains viable, the Golden Ratio solar angle prediction is also disfavoured. This provides a motivation for considering the Golden Ratio predictions 
at NLO. Unlike the case of TB mixing, however, where the LO solar angle prediction is viable, and so one may preserve a subgroup of the Klein symmetry, in the case of the Golden Ratio predictions there is a strong motivation to completely break the original Klein symmetry at the NLO.

In this paper we have proposed a new $A_5$ model of leptons which corrects the LO predictions
of Golden Ratio mixing via a minimal NLO Majorana mass correction which completely breaks the original
Klein symmetry of the neutrino mass matrix. The minimal
nature of the NLO correction leads to a restricted and correlated range of the mixing angles
allowing agreement within the one sigma range of recent global fits.  Yet even though the Golden Model cannot predict the sign of the correction to the solar and atmospheric mixing angles, agreement with recent global fits can be reached in a specific region of parameter space allowed by the model. Remarkably, 
the minimal NLO correction also preserves the LO inverse neutrino mass sum rule, 
implying a neutrino mass spectrum which extends into the quasi-degenerate region,
allowing the model to be accessible to the current and future neutrinoless double beta decay experiments.

\section*{Acknowledgments}
We would like to thank C. Luhn, A. Lytle, A. Merle for useful discussions, and A. Merle for his code for generating $m_{\beta\beta}$ vs. $m_{lightest}$.   The authors acknowledge partial support from the  European Union FP7  ITN INVISIBLES (Marie Curie Actions, PITN- GA-2011- 289442),
and EU ITN grant UNILHC 237920 (Unification in the LHC era), as well as the 
STFC Consolidated ST/J000396/1 .
\newpage

\appendix\numberwithin{equation}{section}
\makeatletter
\def\@seccntformat#1{Appendix\ \csname the#1\endcsname:\;}
\makeatother
\section{The Group Theory of $A_5$}
\label{sec:appendixa}
$A_5$ is a \textit{simple} group.\footnote{$A_5$ contains no nontrivial normal subgroups.}  By definition, it is the group of all even permutations of a set of five elements.  As such, it has $5!/2=60$ elements.  In fact, it can be shown to be isomorphic to the Icosahedral Symmetry Group, $\mathcal{I}$, the group of all rotations of an icosahedron which preserve the icosahedron's orientation.  This isomorphism turns out to be insightful when considering the conjugacy classes of $A_5\cong \mathcal{I}$, as one can express the conjugacy classes in terms of Schoenflies notation where they are denoted by $C_n^k$ and represent rotations by $\frac{2\pi k}{n}$.  With the further definition that the number in front of $C_n^k$ is the number of elements contained in a conjugacy class, the conjugacy classes of $A_5$ can be written as: $I$ (the conjugacy class consisting solely of the identity), $15C^1_2$, $20C^1_3$, $12 C^1_5$ and $12 C_5^2$.  Using these conjugacy classes and the theorems that posit that the number of elements of a group is equal to the sum of the squares of the dimensions of the irreducible representations and that the number of irreducible representations of a group is equal to the number of conjugacy classes, it is easy to see that $A_5$ has a one dimensional irreducible representation ($\bf{1}$), two three dimensional irreducible representations ($\bf{3}$ and $\bf{3^{\prime}}$), one four dimensional representation ($\bf{4}$), and one five dimensional irreducible representation ($\bf{5}$) because
\begin{equation}
1+15+20+12+12=60=1^2+3^2+3^2+4^2+5^2.
\end{equation}
Yet in order to do anything physically useful with these five irreducible representations, an explicit matrix representation for each irreducible representation of $A_5$ must be calculated/found.  To do this it is helpful to find a set of generators and rules defining their multiplication (\textit{i.e.} a presentation) which generate $A_5$.  We choose to work in the basis given in Ref. \cite{goldending} in which $A_5$ is generated by two generators, $S$ and $T$, satisfying the presentation rules
\begin{equation}
S^2=T^5=(ST)^3=1.
\end{equation}
The explicit forms of $S$ and $T$ can be found in Ref. \cite{goldending}.  For completeness, we choose to list them again here along with another element of $A_5$, $U=T^3ST^2ST^3S$.  Notice that the $S$ and $U$ elements generate a $Z_2^S\times Z_2^U$ Klein subgroup of $A_5$.  Recall that in Section \ref{sec:LOModel} of this work, spontaneously breaking $A_5$ to this Klein subgroup (in the neutrino sector) was used to generate Golden Ratio mixing.  The explicit forms of $S$, $T$ and $U$ for each irreducible representation can be found in Table \ref{tab:A5Gen}.

\begin{table}
\begin{eqnarray}\nonumber \footnotesize
\label{A5gen}
\begin{array}{|c|c|c|c|}\hline
&&&\\
&S&T&U\\&&&\\\hline
&&&\\
\bf{1}:&1&1&1\\&&&\\\hline
&&&\\
\bf{3}:&\frac{1}{\sqrt{5}}\left(
\begin{array}{ccc}
 1 & -\sqrt{2} & -\sqrt{2} \\
 -\sqrt{2} & -\phi_g  & \frac{1}{\phi_g } \\
 -\sqrt{2} & \frac{1}{\phi_g } & -\phi_g
\end{array}
\right)&\left(
\begin{array}{ccc}
 1 & 0 & 0 \\
 0 & \rho  & 0 \\
 0 & 0 & \rho ^4
\end{array}
\right)&\left(
\begin{array}{ccc}
 -1 & 0 & 0 \\
 0 & 0 & -1 \\
 0 & -1 & 0
\end{array}
\right)\\&&&\\\hline
&&&\\
\bf{3^{\prime}}:&\frac{1}{\sqrt{5}}\left(
\begin{array}{ccc}
 -1 & \sqrt{2} & \sqrt{2} \\
 \sqrt{2} & -\frac{1}{\phi_g } & \phi_g  \\
 \sqrt{2} & \phi_g  & -\frac{1}{\phi_g }
\end{array}
\right)&\left(
\begin{array}{ccc}
 1 & 0 & 0 \\
 0 & \rho ^2 & 0 \\
 0 & 0 & \rho ^3
\end{array}
\right)&\left(
\begin{array}{ccc}
 -1 & 0 & 0 \\
 0 & 0 & -1 \\
 0 & -1 & 0
\end{array}
\right)\\&&&\\\hline
&&&\\
\bf{4}:&\frac{1}{\sqrt{5}}\left(
\begin{array}{cccc}
 1 & \frac{1}{\phi_g } & \phi_g  & -1 \\
 \frac{1}{\phi_g } & -1 & 1 & \phi_g  \\
 \phi_g  & 1 & -1 & \frac{1}{\phi_g } \\
 -1 & \phi_g  & \frac{1}{\phi_g } & 1
\end{array}
\right)&\left(
\begin{array}{cccc}
 \rho  & 0 & 0 & 0 \\
 0 & \rho ^2 & 0 & 0 \\
 0 & 0 & \rho ^3 & 0 \\
 0 & 0 & 0 & \rho ^4
\end{array}
\right)&\left(
\begin{array}{cccc}
 0 & 0 & 0 & 1 \\
 0 & 0 & 1 & 0 \\
 0 & 1 & 0 & 0 \\
 1 & 0 & 0 & 0
\end{array}
\right)\\&&&\\\hline
&&&\\
\bf{5}:&\frac{1}{5} \left(
\begin{array}{ccccc}
 -1 & \sqrt{6} & \sqrt{6} & \sqrt{6} & \sqrt{6} \\
 \sqrt{6} & \frac{1}{\phi_g ^2} & -2 \phi_g  & \frac{2}{\phi_g } & \phi_g ^2 \\
 \sqrt{6} & -2 \phi_g  & \phi_g ^2 & \frac{1}{\phi_g ^2} & \frac{2}{\phi_g } \\
 \sqrt{6} & \frac{2}{\phi_g } & \frac{1}{\phi_g ^2} & \phi_g ^2 & -2 \phi_g  \\
 \sqrt{6} & \phi_g ^2 & \frac{2}{\phi_g } & -2 \phi_g  & \frac{1}{\phi_g ^2}
\end{array}
\right)&\left(
\begin{array}{ccccc}
 1 & 0 & 0 & 0 & 0 \\
 0 & \rho  & 0 & 0 & 0 \\
 0 & 0 & \rho ^2 & 0 & 0 \\
 0 & 0 & 0 & \rho ^3 & 0 \\
 0 & 0 & 0 & 0 & \rho ^4
\end{array}
\right)&\left(
\begin{array}{ccccc}
 1 & 0 & 0 & 0 & 0 \\
 0 & 0 & 0 & 0 & 1 \\
 0 & 0 & 0 & 1 & 0 \\
 0 & 0 & 1 & 0 & 0 \\
 0 & 1 & 0 & 0 & 0
\end{array}
\right)\\
&&&\\\hline
\end{array}
\end{eqnarray}
\caption{The $S$, $T$, and $U$ elements of $A_5$ for the $\bf{1}$-, $\bf{3}$-, $\bf{3^{\prime}}$-, $\bf{4}$-, and $\bf{5}$-dimensional irreducible representations in which $\rho=e^{\frac{2\pi i}{5}}$ and $\phi_g=(1+\sqrt{5})/2$ is the Golden Ratio.}
\label{tab:A5Gen}
\end{table}
Each of the elements in Table \ref{tab:A5Gen} will leave invariant a different flavon VEV alignment (up to proportionality), \textit{i.e.} $G\langle\phi\rangle=\langle\phi\rangle$ where $G=S,T$ or $U$ and $\langle\phi\rangle$ is the VEV of a flavon field preserved by the action of the group element $G$.  These alignments are crucial when considering the spontaneous breaking of $A_5$ to $Z_2^S\times Z_2^U$, as having flavons develop VEVs invariant under the action of the $S$ and $U$ preserves the Klein subgroup.  A listing of these ``invariant'' alignments is given in Table \ref{tab:invvev}.

\begin{table}[ht!]\centering
\begin{tabular}{|c|c|c|c|}\hline
&&&\\
&$S$&$T$&$U$\\&&&\\\hline&&&\\
\bf{$3$}& 
$\left(
\begin{array}{c}
 1 \\
 -\frac{1}{\sqrt{2} \phi_g } \\
 -\frac{1}{\sqrt{2} \phi_g }
\end{array}
\right)$&$\left(
\begin{array}{c}
 1 \\
 0 \\
 0
\end{array}
\right)$&$\left(
\begin{array}{c}
 0 \\
 1 \\
 -1
\end{array}
\right)$\\&&&\\\hline&&&\\
\bf{$3^{\prime}$} &$\left(
\begin{array}{c}
 1 \\
 \frac{\phi_g }{\sqrt{2}} \\
 \frac{\phi_g }{\sqrt{2}}
\end{array}
\right)$& $\left(
\begin{array}{c}
 1 \\
 0 \\
 0
\end{array}
\right)$&$\left(
\begin{array}{c}
 0 \\
 1 \\
 -1
\end{array}
\right)$\\&&&\\\hline &&&\\
\bf{$4$}&
$\left(
\begin{array}{c}
 w_3 \phi_g ^2-w_2 \phi_g  \\
 w_2 \\
 w_3 \\
 w_2 \phi_g ^2-w_3 \phi_g 
\end{array}
\right)$&-&$\left(
\begin{array}{c}
 w_1 \\
 w_2 \\
 w_2 \\
 w_1
\end{array}
\right)$\\&&&\\\hline&&&\\
\bf{$5$}&$\left(
\begin{array}{c}
 \frac{v_4}{\sqrt{6} \phi_g ^2}+\frac{v_3 \phi_g }{\sqrt{6}}+\sqrt{\frac{2}{3}} v_2 \\
 v_2 \\
 v_3 \\
 v_4 \\
 \frac{v_3}{\phi_g }-\frac{v_4}{\phi_g }+v_2
\end{array}
\right)$&$\left(
\begin{array}{c}
 1 \\
 0 \\
 0 \\
 0 \\
 0
\end{array}
\right)$&$\left(
\begin{array}{c}
 v_1 \\
 v_2 \\
 v_3 \\
 v_3 \\
 v_2
\end{array}
\right)$\\&&&\\\hline
\end{tabular}
\caption{The Invariant VEV Alignments for one and only one of the $S$, $T$, and $U$ elements.  The $w_i$ and $v_i$ represent components of the $\bf{4}$- and $\bf{5}$-dimensional flavon VEVs left unspecified by $S$ and $U$, respectively.  Notice that there exists no nontrivial VEV invariant under the action of $T_4$.}
\label{tab:invvev}
\end{table}
Notice that if \textit{both} $S$ \textit{and} $U$ generators are unbroken, such that the Klein symmetry exists at low energies, then the only nontrivial alignments that exist are for the $\bf{4}$- and $\bf{5}$-dimensional irreducible representations.  The VEV which preserves both $S_4$ \textit{and} $U_4$ is given by
\begin{eqnarray}
\left(
\begin{array}{c}
 1 \\
 1 \\
 1 \\
 1
\end{array}
\right),
\end{eqnarray}
whereas the VEV which preserves $S_5$ \textit{and} $U_5$ is given by
\begin{equation}
 \begin{pmatrix}
   \sqrt{\frac{2}{3}}\left(v_2+v_3\right) \\
   v_2 \\
   v_3 \\
   v_3 \\
   v_2
 \end{pmatrix}.
\end{equation}
Recall that the five dimensional VEV given above was crucial in constructing the LO Golden Model of Section \ref{sec:LOModel}.
In addition to constructing these invariant VEVs the $S$ and $T$ elements/generators can also be used to calculate the Clebsch-Gordan coefficients associated with the particular irreducible representations of $A_5$.  This is the goal of the remainder of this appendix.

\newpage

\subsection{The Kronecker Products and Clebsch-Gordan Coefficients of $A_5$}

From the matrix representations of the $S$ and $T$ elements/generators of $A_5$ given in the previous section, it is straightforward to calculate the Clebsch-Gordan coefficients for the decomposition of the product representations, which we now list for this basis in detail.  We use $a_i$ to denote the elements of the first representation, $b_i$ to indicate those of the second representation of the product, and the subscripts ``$\bf{a}$" and ``$\bf{s}$" to indicate a representation which is antisymmetric or symmetric, respectively.\\
~\newline

\begin{eqnarray}
\begin{array}{|c|c|}\hline
\fbox{$\mathbf{3}\otimes\mathbf{3}=\mathbf{1_s}\oplus\mathbf{3_a} \oplus\mathbf{5_s}$}&\fbox{$\mathbf{3}^{\prime}\otimes\mathbf{3}'=\mathbf{1_s}\oplus\mathbf{3'_a}\oplus\mathbf{5_s}$}\\
&\\
\mathbf{1_s}\sim a_1 b_1+a_2 b_3+a_3 b_2&\mathbf{1_s}\sim a_1 b_1+a_2 b_3+a_3 b_2\\
&\\
\nonumber\mathbf{3_a}\sim\left(
\begin{array}{c}
 a_2 b_3-a_3 b_2 \\
 a_1 b_2-a_2 b_1 \\
 a_3 b_1-a_1 b_3
\end{array}
\right)&\mathbf{3'_a}\sim\left(
\begin{array}{c}
 a_2 b_3-a_3 b_2 \\
 a_1 b_2-a_2 b_1 \\
 a_3 b_1-a_1 b_3
\end{array}
\right)\\
&\\
\mathbf{5_s}\sim\left(
\begin{array}{c}
 2 a_1 b_1-a_2 b_3-a_3 b_2 \\
 -\sqrt{3} a_1 b_2-\sqrt{3} a_2 b_1 \\
 \sqrt{6} a_2 b_2 \\
 \sqrt{6} a_3 b_3 \\
-\sqrt{3} a_1 b_3-\sqrt{3} a_3 b_1
\end{array}
\right)&\mathbf{5_s}\sim\left(
\begin{array}{c}
 2 a_1 b_1-a_2 b_3 -a_3 b_2\\
 \sqrt{6} a_3 b_3 \\
 -\sqrt{3} a_1 b_2 -\sqrt{3} a_2 b_1\\
 -\sqrt{3} a_1 b_3-\sqrt{3} a_3 b_1 \\
 \sqrt{6} a_2 b_2
\end{array}\right)\\
&\\\hline
\end{array}
\end{eqnarray}

\hspace{.35in}\begin{centering}
\fbox{
\begin{minipage}[h]{4.451 in}
\centering{
\fbox{$\mathbf{3}\otimes\mathbf{3}'=\mathbf{4}\oplus\mathbf{5}$}}
\begin{eqnarray}
\begin{array}{cc}
\nonumber\mathbf{4}\sim\left(
\begin{array}{c}
 \sqrt{2} a_2 b_1+a_3 b_2 \\
 -\sqrt{2} a_1 b_2-a_3 b_3 \\
 -\sqrt{2} a_1 b_3-a_2 b_2 \\
 \sqrt{2} a_3 b_1+a_2 b_3
\end{array}
\right)&
\mathbf{5}\sim\left(
\begin{array}{c}
 \sqrt{3} a_1 b_1 \\
 a_2 b_1-\sqrt{2} a_3 b_2 \\
 a_1 b_2-\sqrt{2} a_3 b_3 \\
 a_1 b_3-\sqrt{2} a_2 b_2 \\
 a_3 b_1-\sqrt{2} a_2 b_3
\end{array}
\right)\\
&\\
\end{array}
\end{eqnarray}
\end{minipage}}
\end{centering}

\newpage

\begin{eqnarray}\nonumber
\begin{array}{|c|c|}\hline
\fbox{$\mathbf{3}\otimes\mathbf{4}=\mathbf{3}'\oplus\mathbf{4}\oplus\mathbf{5}$}&\fbox{$\mathbf{3}'\otimes\mathbf{4}=\mathbf{3}\oplus\mathbf{4}\oplus\mathbf{5}$}\\
&\\
\mathbf{3}'\sim\left(
\begin{array}{c}
 -\sqrt{2} a_2 b_4-\sqrt{2} a_3 b_1 \\
 \sqrt{2} a_1 b_2-a_2 b_1+a_3 b_3 \\
 \sqrt{2} a_1 b_3+a_2 b_2-a_3 b_4
\end{array}
\right)&\mathbf{3}\sim\left(
\begin{array}{c}
 -\sqrt{2} a_2 b_3-\sqrt{2} a_3 b_2 \\
 \sqrt{2} a_1 b_1+a_2 b_4-a_3 b_3 \\
  \sqrt{2} a_1 b_4 -a_2 b_2+a_3 b_1
\end{array}
\right)\\
&\\
\mathbf{4}\sim\left(
\begin{array}{c}
  a_1 b_1-\sqrt{2}a_3 b_2 \\
 -a_1 b_2-\sqrt{2}a_2 b_1 \\
  a_1 b_3+\sqrt{2}a_3 b_4 \\
  -a_1 b_4+\sqrt{2}a_2 b_3
\end{array}
\right)&\mathbf{4}\sim\left(
\begin{array}{c}
 a_1 b_1+\sqrt{2}a_3 b_3 \\
  a_1 b_2-\sqrt{2}a_3 b_4 \\
 -a_1 b_3+\sqrt{2}a_2 b_1 \\
   -a_1 b_4-\sqrt{2}a_2 b_2
\end{array}
\right)\\
&\\
\mathbf{5}\sim\left(
\begin{array}{c}
 \sqrt{6} a_2
   b_4-\sqrt{6} a_3 b_1 \\
 2\sqrt{2} a_1 b_1+2 a_3 b_2\\
 -\sqrt{2} a_1 b_2+a_2 b_1+3a_3 b_3 \\
\sqrt{2} a_1 b_3-3a_2 b_2-a_3 b_4\\
 -2\sqrt{2} a_1 b_4-2 a_2 b_3
\end{array}
\right)&\mathbf{5}\sim\left(
\begin{array}{c}
 \sqrt{6} a_2 b_3-\sqrt{6} a_3
   b_2 \\
   \sqrt{2} a_1 b_1-3a_2 b_4-a_3 b_3 \\
   2\sqrt{2} a_1 b_2+2 a_3 b_4
   \\
   -2\sqrt{2} a_1 b_3-2a_2 b_1 \\
 -\sqrt{2}a_1 b_4+a_2 b_2+3a_3 b_1
\end{array}
\right)\\
&\\\hline
\fbox{$\mathbf{3}\otimes\mathbf{5}=\mathbf{3}\oplus\mathbf{3}'\oplus\mathbf{4}\oplus\mathbf{5}$}&\fbox{$\mathbf{3}'\otimes\mathbf{5}=\mathbf{3}\oplus\mathbf{3}'\oplus\mathbf{4}\oplus\mathbf{5}$}\\
&\\
\mathbf{3}\sim\left(
\begin{array}{c}
 -2 a_1 b_1+\sqrt{3}a_2 b_5+\sqrt{3}a_3 b
   _2 \\
 \sqrt{3}a_1 b
   _2+a_2 b_1-\sqrt{6}a_3 b_3 \\
 \sqrt{3}a
   _1 b_5-\sqrt{6}a_2 b_4+a_3 b_1
\end{array}
\right)&\mathbf{3}\sim\left(
\begin{array}{c}
 \sqrt{3} a_1 b
   _1+a_2b_4+a_3 b_3 \\
 a_1 b_2-\sqrt{2}
   a_2 b_5 -\sqrt{2}
   a_3 b_4\\
 a_1 b_5-\sqrt{2} a_2 b_3-\sqrt{2} a_3 b
   _2
\end{array}
\right)\\
&\\
\mathbf{3}'\sim\left(
\begin{array}{c}
 \sqrt{3} a_1 b_1+a_2 b_5+a_3 b
   _2 \\
  a_1 b_3-\sqrt{2}a_2 b_2-\sqrt{2}a_3 b
   _4 \\
 a_1 b_4-\sqrt{2}a_2 b_3-\sqrt{2}a_3 b
   _5
\end{array}
\right)&\mathbf{3}'\sim\left(
\begin{array}{c}
 -2 a_1 b_1+\sqrt{3}
   a_2 b_4 +\sqrt{3}
   a_3 b_3\\
 \sqrt{3}
   a_1 b_3+a_2 b_1-\sqrt{6}
   a_3 b_5 \\
 \sqrt{3}
   a_1 b_4-\sqrt{6}
   a_2 b_2+a_3 b_1
\end{array}
\right)\\
&\\
\mathbf{4}\sim\left(
\begin{array}{c}
 2\sqrt{2} a_1 b
   _2-\sqrt{6} a_2 b_1+a_3 b_3 \\
 -\sqrt{2}a_1 b_3+2a_2 b_2-3 a
   _3 b_4 \\
 \sqrt{2}a_1 b
   _4+3a_2 b_3-2a_3 b_5 \\
 -2\sqrt{2} a_1 b_5-a_2 b
   _4+\sqrt{6} a_3 b_1
\end{array}
\right)&\mathbf{4}\sim\left(
\begin{array}{c}
 \sqrt{2} a_1 b_2+3 a_2
   b_5-2a_3 b_4 \\
 2\sqrt{2} a_1 b_3-\sqrt{6} a_2 b_1+a_3 b_5 \\
 -2\sqrt{2} a_1 b_4-a_2 b_2 +\sqrt{6} a_3 b
   _1\\
 -\sqrt{2} a_1b_5+2 a_2
   b_3-3 a_3 b_2
\end{array}
\right)\\
&\\
\mathbf{5}\sim\left(
\begin{array}{c}
 \sqrt{3} a_2 b_5-\sqrt{3} a_3
   b_2 \\
 -a_1 b_2-\sqrt{3} a_2 b_1-\sqrt{2}a_3 b_3 \\
 -2 a_1 b_3-\sqrt{2}a_2 b_2 \\
 2a_1 b_4+\sqrt{2}a_3 b_5 \\
 a_1 b_5+\sqrt{2}a_2 b_4+ \sqrt{3} a_3 b_1
\end{array}
\right)&\mathbf{5}\sim\left(
\begin{array}{c}
 \sqrt{3} a_2 b
   _4-\sqrt{3} a_3 b_3
   \\
 2 a_1 b_2+\sqrt{2}
   a_3 b_4 \\
 -a_1 b_3-\sqrt{3} a_2 b
   _1-\sqrt{2}a_3 b_5 \\
a_1 b_4+\sqrt{2} a_2 b_2 + \sqrt{3} a_3 b
   _1\\
 -2a_1 b_5-\sqrt{2} a_2 b_3
\end{array}
\right)\\
&\\\hline
\end{array}
\end{eqnarray}
\newpage
~\newline
\newline

\begin{eqnarray}\nonumber\small
\begin{array}{|c|c|}\hline
\fbox{$\mathbf{4}\otimes\mathbf{4}=\mathbf{1_s}\oplus\mathbf{3_a}\oplus\mathbf{3'_a}\oplus\mathbf{4_s}\oplus\mathbf{5_s}$}&\fbox{$\mathbf{4}\otimes\mathbf{5}=\mathbf{3}\oplus\mathbf{3}'\oplus\mathbf{4}\oplus\mathbf{5}_1\oplus\mathbf{5}_2$}\\
&\\
\mathbf{1_s}\sim a_1b_4+a_2 b_3+a_3 b_2+a_4 b_1&\mathbf{3}\sim\left(
\begin{array}{c}
 2 \sqrt{2} a_1b_5-\sqrt{2} a_2 b_4+\sqrt{2} a_3 b_3-2 \sqrt{2} a_4 b_2\\
 -\sqrt{6} a_1 b_1+2 a_2 b_5+3 a_3 b_4-a_4 b_3 \\
 a_1 b_4-3 a_2b_3-2a_3 b_2+\sqrt{6} a_4 b_1
\end{array}
\right)\\
&\\
\nonumber\mathbf{3_a}\sim\left(
\begin{array}{c}
 -a_1 b_4+a_2b_3-a_3
   b_2+a_4 b_1\\
 \sqrt{2} a_2 b
   _4-\sqrt{2} a_4 b_2
   \\
 \sqrt{2} a_1 b
   _3-\sqrt{2} a_3 b_1
\end{array}
\right)&\mathbf{3}'\sim\left(
\begin{array}{c}
 \sqrt{2} a_1 b_5+2\sqrt{2} a_2 b_4-2\sqrt{2} a_3 b_3-\sqrt{2} a_4 b_2   \\
 3a_1 b_2-\sqrt{6} a_2 b_1-a_3 b_5+2 a_4b_4 \\
 -2 a_1 b_3+a_2 b_2+\sqrt{6} a_3 b_1-3 a_4 b
   _5
\end{array}
\right)\\
&\\
\mathbf{3'_a}\sim\left(
\begin{array}{c}
a_1 b_4 +a_2 b_3-a_3
   b_2 -a_4 b_1\\
 \sqrt{2} a_3 b
   _4-\sqrt{2} a_4 b_3
   \\
 \sqrt{2} a_1 b
   _2-\sqrt{2} a_2 b_1
\end{array}
\right)&\mathbf{4}\sim\left(
\begin{array}{c}
 \sqrt{3} a_1 b_1-\sqrt{2} a_2 b_5+\sqrt{2} a_3 b_4-2\sqrt{2} a_4 b_3
   \\
 -\sqrt{2} a_1 b_2-\sqrt{3} a_2 b_1+2 \sqrt{2} a_3 b_5+\sqrt{2} a_4 b_4 \\
 \sqrt{2} a_1 b_3+2\sqrt{2} a_2 b_2-\sqrt{3} a_3 b_1-\sqrt{2} a_4 b_5
   \\
 -2 \sqrt{2} a_1 b_4+\sqrt{2} a_2 b_3-\sqrt{2} a_3 b_2+\sqrt{3} a_4 b_1
\end{array}
\right)\\&\\
\mathbf{4_s}\sim\left(
\begin{array}{c}
 a_2 b_4+a_3b_3+a_4 b_2 \\
 a_1 b_1+a_3 b_4 +a_4 b_3\\
 a_1b_2+a_2 b_1+a_4 b_4 \\
 a_1 b_3+a_2b_2+a_3 b_1
\end{array}
\right)&\mathbf{5}_1\sim\left(
\begin{array}{c}
 \sqrt{2} a_1 b_5-\sqrt{2} a_2 b_4-\sqrt{2} a_3 b_3+\sqrt{2} a_4 b_2
   \\
 -\sqrt{2} a_1 b_1-\sqrt{3} a_3 b_4 -\sqrt{3} a_4 b_3\\
 \sqrt{3} a_1 b_2+\sqrt{2} a_2 b_1+\sqrt{3} a_3 b_5
   \\
 \sqrt{3} a_2 b_2+\sqrt{2} a_3 b_1+\sqrt{3} a_4 b_5
   \\
 -\sqrt{3} a_1 b_4-\sqrt{3} a_2 b_3-\sqrt{2} a_4 b_1
\end{array}
\right)\\
&\\
\mathbf{5_s}\sim\left(
\begin{array}{c}
 \sqrt{3} a_1 b_4-\sqrt{3} a_2 b
   _3-\sqrt{3} a_3 b_2+\sqrt{3} a_4 b_1
   \\
 -\sqrt{2} a_2 b_4+2
   \sqrt{2} a_3 b_3-\sqrt{2} a_4 b_2
   \\
 -2 \sqrt{2} a_1 b
   _1+\sqrt{2} a_3 b_4+\sqrt{2} a_4 b
   _3
   \\
 \sqrt{2} a_1 b
   _2+\sqrt{2} a_2 b
   _1-2 \sqrt{2} a_4 b
   _4 \\
 -\sqrt{2} a_1 b_3+2\sqrt{2} a_2 b_2-\sqrt{2} a_3 b_1
\end{array}
\right)&\mathbf{5}_2\sim\left(
\begin{array}{c}
 2 a_1b_5+4 a_2 b_4+4 a_3 b_3 +2 a_4 b_2\\
 4 a_1 b_1+2 \sqrt{6} a_2 b_5 \\
 -\sqrt{6} a_1b_2+2 a_2 b_1-\sqrt{6} a_3 b_5 +2 \sqrt{6} a_4 b_4\\
 2 \sqrt{6} a_1 b_3-\sqrt{6} a_2b_2+2 a_3 b_1-\sqrt{6} a_4 b_5 \\
 2 \sqrt{6} a_3 b_2+4 a_4 b_1
\end{array}
\right)\\
&\\\hline
\end{array}
\end{eqnarray}

\newpage

\begin{centering}
\fbox{
\begin{minipage}[h]{5.5 in}
\hspace{1.2in}
\fbox{$\mathbf{5}\otimes\mathbf{5}=\mathbf{1_s}\oplus\mathbf{3_a}\oplus\mathbf{3'_a}\oplus\mathbf{4_s}\oplus\mathbf{4_a}\oplus\mathbf{5_{1,s}}\oplus\mathbf{5_{2,s}}$}
\begin{eqnarray}\nonumber
\begin{array}{c}
\mathbf{1}_s\sim a_1b_1+a_2b_5+a_3b_4+a_4b_3+a_5b_2\\\\
\mathbf{3}_a\sim\left(
\begin{array}{c}
 a_2 b_5+2a_3 b_4-2 a_4 b_3-a_5 b_2 \\
 -\sqrt{3} a_1b_2+\sqrt{3} a_2 b_1+\sqrt{2}a_3 b_5-\sqrt{2} a_5 b_3 \\
 \sqrt{3}a_1 b_5+\sqrt{2} a_2 b_4-\sqrt{2} a_4b_2-\sqrt{3} a_5 b_1
\end{array}
\right)\\\\
\mathbf{3'_a}\sim\left(
\begin{array}{c}
 2 a_2 b_5-a_3 b_4+a_4 b_3-2 a_5 b_2\\
 \sqrt{3} a_1b_3-\sqrt{3} a_3 b_1+\sqrt{2}a_4 b_5-\sqrt{2} a_5 b_4 \\
 -\sqrt{3}a_1 b_4+\sqrt{2} a_2 b_3-\sqrt{2} a_3b_2+\sqrt{3} a_4 b_1
\end{array}
\right)\\\\
\mathbf{4_s}\sim\left(
\begin{array}{c}
 3 \sqrt{2} a_1 b_2+3 \sqrt{2} a_2 b_1-\sqrt{3}a_3b_5+4 \sqrt{3} a_4 b_4-\sqrt{3}a_5 b_3 \\
 3 \sqrt{2}a_1 b_3+4 \sqrt{3} a_2 b_2+3 \sqrt{2} a_3 b_1-\sqrt{3}a_4b_5-\sqrt{3} a_5 b_4 \\
 3 \sqrt{2} a_1 b_4-\sqrt{3}a_2 b_3-\sqrt{3} a_3 b_2+3 \sqrt{2} a_4 b_1+4 \sqrt{3}a_5b_5 \\
 3 \sqrt{2} a_1b_5-\sqrt{3} a_2 b_4+4 \sqrt{3}a_3 b_3-\sqrt{3} a_4 b_2+3 \sqrt{2} a_5 b_1
\end{array}
\right)\\\\

\mathbf{4_a}\sim\left(
\begin{array}{c}
 \sqrt{2} a_1 b_2-\sqrt{2} a_2 b_1+\sqrt{3} a_3 b_5-\sqrt{3}a_5 b_3 \\
 -\sqrt{2} a_1 b_3+\sqrt{2} a_3 b_1+\sqrt{3} a_4 b_5-\sqrt{3} a_5 b_4 \\
 -\sqrt{2} a_1 b_4-\sqrt{3} a_2 b_3+\sqrt{3} a_3 b_2+\sqrt{2} a_4 b_1\\
 \sqrt{2} a_1 b_5-\sqrt{3}a_2 b_4+\sqrt{3} a_4 b_2-\sqrt{2} a_5 b_1
\end{array}
\right)\\\\

\mathbf{5_{1,s}}\sim\left(
\begin{array}{c}
 2 a_1 b_1+a_2 b_5-2 a_3 b_4-2 a_4 b_3+a_5 b_2 \\
 a_1 b_2+a_2 b_1+\sqrt{6} a_3 b_5+\sqrt{6} a_5 b_3 \\
 -2 a_1 b_3+\sqrt{6} a_2 b_2-2 a_3 b_1 \\
 -2 a_1 b_4-2 a_4 b_1+\sqrt{6} a_5 b_5 \\
 a_1 b_5+\sqrt{6} a_2 b_4+\sqrt{6} a_4 b_2+a_5 b_1
\end{array}
\right)\\\\

\mathbf{5_{2,s}}\sim\left(
\begin{array}{c}
 2 a_1 b_1-2 a_2 b_5+a_3 b_4+a_4b_3-2 a_5 b_2 \\
 -2 a_1 b_2-2 a_2 b_1+\sqrt{6} a_4 b_4 \\
 a_1 b_3+a_3 b_1+\sqrt{6} a_4 b_5+\sqrt{6} a_5 b_4 \\
 a_1 b_4+\sqrt{6} a_2 b_3+\sqrt{6} a_3 b_2+a_4 b_1 \\
 -2 a_1 b_5+\sqrt{6} a_3 b_3-2 a_5 b_1
 \end{array}
\right)\\\\
\end{array}
\end{eqnarray}
\end{minipage}}
\end{centering}

\newpage

\section{Breaking the GR Klein Symmetry}
\label{sec:appendixc}


In this appendix we show that the assumed correction to the superpotential (cf. Eq. (\ref{eqn:newMaj})),
\begin{equation}
 \Delta W_{\nu}=\frac{y_{A}}{\Lambda}\left(NN\right)_{\bf 1_s}\left(\lambda\lambda\right)_{\bf 1_s}+\frac{y_{C}}{\Lambda}\left(\left(NN\right)_{\bf 5_s}\left(\lambda\lambda\right)_{\bf 5_s}\right)_{\bf 1_s},
\end{equation}
 is sufficient to completely break the Klein symmetry associated with the unsuccessful LO Golden ratio prediction.
We begin by first noting that so long as $a_1\neq 0$ and/or $a_2\neq-a_3$, a non-zero value of $\theta_{13}$ will be obtained because a VEV proportional to $(0,1,-1)^T$ will be preserved by $U_{3}$ (cf. Table \ref{tab:invvev}).
Furthermore, by construction, any representation formed from the combination of two triplet fields preserving $U_3$ will preserve said representation's corresponding $U$.  For example, the $\bf{5}$ constructed from two $\lambda$ fields where the $a_i$ are left arbitrary is 
\begin{eqnarray}
\langle(\lambda\lambda)_{\bf{5_s}}\rangle\sim
\left(
\begin{array}{c}
 2 a_1^2-2 a_2 a_3 \\
 -2 \sqrt{3} a_1 a_2 \\
 \sqrt{6} a_2^2 \\
 \sqrt{6} a_3^2 \\
 -2 \sqrt{3} a_1 a_3
\end{array}
\right).
\end{eqnarray}
In the above result, the subscript ``$\bf{5_s}$'' on $(\lambda\lambda)$ denotes that we are selecting the $\bf{5_s}$ from the tensor product $\bf{3\otimes 3=1_s\oplus 3_a\oplus 5_s}$.  From the form of $U_{5}$ given in Table \ref{tab:A5Gen}, 
it is clear to see that only two solutions exist which leave $U_5$ invariant (\textit{i.e.} satisfy $U_5\langle\left(\lambda\lambda\right)_{\bf 5_s}\rangle=\langle\left(\lambda\lambda\right)_{\bf 5_s}\rangle$):
\begin{eqnarray}
\left\{\begin{array}{cc}
a_1=0~\text{and}~a_2=-a_3\\
a_2=a_3~\text{and}~a_1~\text{arbitrary}
\end{array}\right.
\end{eqnarray}
From these two solutions, the latter may be interpreted as resulting from general invariance of a VEV acted upon by $U_{5}$ and the former existing because the $\bf{5_s}$ was constructed from \textit{two} $\bf{3}$- dimensional irreducible representations.  In either case,  the assumed VEV of $\vev{\lambda}=(0,0,a_3)^T$ breaks $U_3$ and the corresponding $\langle\left(\lambda\lambda\right)_{\bf 5_s}\rangle$ breaks $U_5$ providing a correction to the problematic vanishing reactor mixing angle.  Next, we turn to the invariance of $\langle\left(\lambda\lambda\right)_{\bf 5_s}\rangle$ under the $S$ generator.  From the form of $S_{3}$ (cf. Table \ref{tab:A5Gen}),

it is straightforward to see a $\vev{\lambda}\propto (-\sqrt{2}\phi_g,1,1)^T$ is left unchanged by the action of $S_3$ (cf. Table \ref{tab:invvev}).  This, in turn, implies the conditions for the invariance of a triplet VEV under the $\bf{3}$- dimensional $S$ generator:
\begin{eqnarray}
\begin{array}{ccc}
a_1=-\sqrt{2}\phi_g a_3&~\text{and}~ & a_2=a_3.
\end{array}
\end{eqnarray}
Then, acting $S_5$ (cf. Table \ref{tab:A5Gen})
on $\langle\left(\lambda\lambda\right)_{\bf 5}\rangle$ implies  $S_5$ is preserved by the $\langle\left(\lambda\lambda\right)_{\bf 5}\rangle$ VEVs constructed from 
\begin{eqnarray}
\vev{\lambda}\propto
\begin{array}{cccc}
\left(
\begin{array}{c}
 -\phi_g\sqrt{2}   \\
1 \\
1
\end{array}
\right),&\left(
\begin{array}{c}
 \sqrt{2} (\phi_g -1) \\
 1 \\
 1
\end{array}
\right),&\left(
\begin{array}{c}
 \phi_g\sqrt{2}   \\
  (2 \phi_g +1) \\
 -1
\end{array}
\right),&\left(
\begin{array}{c}
 \frac{a_3+1}{\sqrt{2} \phi_g } \\
 1 \\
 a_3
\end{array}
\right).
\end{array}
\end{eqnarray}
Notice that the first two solutions preserve $S_3$ and $S_{3^{\prime}}$, respectively.  The final two alignments preserve $S_5$ (when it acts on their corresponding $\langle\left(\lambda\lambda\right)_{\bf 5}\rangle$ contraction).  From these $S$-preserving alignments it is easy to see $\vev{\lambda}=(0,0,a_3)^T$ will also break the $S$ generator.  Furthermore, it is interesting to see that,  as noted in Ref. \cite{goldending}, it is impossible to preserve both $S_3$ and $U_3$ simultaneously by any VEV given to a flavon field transforming under the $\bf{3}$-dimensional irreducible representation.  However, this is exactly what is needed to correct the problematic predictions of LO golden ratio mixing given in Eq. (\ref{UGR}).

\pagebreak



\end{document}